\PassOptionsToPackage{comma,numbers,sort&compress,super}{natbib}  
\documentclass[
 reprint,
 amsmath,amssymb,
 prx,
 aps,
 nofootinbib,
 longbibliography,
]{revtex4-2}

\usepackage{graphicx}
\usepackage{mciteplus}
\usepackage{siunitx}
\DeclareSIUnit\angstrom{\protect \text {Å}}
\usepackage{floatflt,epsfig} 
\usepackage{float}
\usepackage{mathtools}
\usepackage{color}
\usepackage{braket}
\usepackage{dcolumn}
\usepackage{hyperref}
\usepackage[normalem]{ulem}
\usepackage{pdfpages}

\usepackage[english]{babel}
\usepackage{blindtext}
\usepackage{lipsum}  
\usepackage{comment}
\usepackage{amsthm}
\usepackage{tabularx}
\usepackage{bm}
\usepackage{dsfont}
\usepackage{bbold}
\usepackage{ulem}
\usepackage{pgffor}
\usepackage{mwe,tikz}
\usepackage[percent]{overpic}
\usepackage{todonotes}
\usepackage{algorithm}
\usepackage{algpseudocode}
\usepackage{uri}

\DeclareUnicodeCharacter{0308}{\"{a}}

\DeclareUnicodeCharacter{0301}{\"{o}}

\setcitestyle{super,open={},close={}}

\usepackage{epstopdf}

\DeclareSIUnit{\atomicunit}{a.u.}
\DeclareSIUnit{\nanometer}{nm}

    \makeatletter
    \def\@fnsymbol#1{\ensuremath{\ifcase#1\or *\or*\or  \ddagger\or
   \mathsection\or \mathparagraph\or \|\or **\or \dagger\dagger
   \or \ddagger\ddagger \else\@ctrerr\fi}}
    \makeatother

\makeatletter
\AtBeginDocument{\let\LS@rot\@undefined}
\makeatother

\begin{document}

\preprint{APS/123-QED}


\title{Realistic ab initio predictions of excimer behavior under \\ collective light-matter strong coupling}

\author{Matteo Castagnola}
\affiliation{Department of Chemistry, Norwegian University of Science and Technology, 7491 Trondheim, Norway}

\author{Marcus T. Lexander}
\affiliation{Department of Chemistry, Norwegian University of Science and Technology, 7491 Trondheim, Norway}

\author{Henrik Koch}
\email{henrik.koch@ntnu.no}
\affiliation{Department of Chemistry, Norwegian University of Science and Technology, 7491 Trondheim, Norway}

\begin{abstract}
Experiments show that light-matter strong coupling affects chemical properties, though the underlying mechanism remains unclear.
We present an \textit{ab initio} quantum electrodynamics coupled cluster method for the collective strong coupling regime. 
The model accurately describes electronic and electron-photon correlation within a molecular subsystem, while a simplified description of the collective polaritonic excitations allows for realistic microscopic light-matter couplings.
We illustrate the model by investigating the potential energy surfaces of the argon dimer.
This provides a prototype for excimers, and we analyze the ground and excited state vibrational levels.
In the collective regime (small light-matter coupling, large number of molecules), the ground state potential energy surface and the first vibrational levels of the excited state are not changed significantly.
However, collective strong coupling produces an abrupt transition in the vibrational landscape of the excimer, causing higher levels to behave similarly to ground state vibrations.
Beyond a critical collective coupling strength, the excimer formation is thus inhibited.
\end{abstract}

\maketitle

\section{Introduction}

Light-matter strong coupling has been shown to modify chemical properties in both the excited (photochemistry) and ground states,\cite{Thomas2019, hutchison2012, Lather2019, zeng2023control, lee2024controlling, hirai2021selective, ahn2023modification} but the mechanism behind these changes is still unclear.
The analysis is complicated by the collective nature of polaritons, which are delocalized over many molecules and show a significant photon component.
For this reason, polaritons are usually studied within the Tavis-Cummings framework,\cite{tavis1968exact, dicke1954coherence} which describes molecules as two-level oscillators and thus cannot model complex chemical processes.
This collective viewpoint conflicts with the chemical intuition that molecular properties are determined by local molecular forces.
For this reason, some \textit{ab initio} methods of quantum chemistry have been extended to include photons explicitly.\cite{haugland2020coupled, ruggenthaler2014quantum, riso2022molecular, mandal2020polarized, bauer2023perturbation}
However, these methods provide a partial view since, due to their steep computational cost, they are limited to a small number of molecules with artificially large light-matter couplings.
There have been significant efforts to develop multiscale models that include several replicas of the reacting molecule (see, e.g., Refs.\cite{groenhof2018coherent, luk2017multiscale, dutta2024thermal, groenhof2019tracking, tichauer2021multi, sokolovskii2024one}).
These models allow for a quantum mechanical molecular description with a reasonably small light-matter coupling strength in the collective regime.
Nevertheless, it is unsatisfying that the details of many molecules in the ensemble are necessary. 
Other methods describe molecules by vibrational levels in a truncated electronic space to determine the collective dynamics of the system,\cite{perez2024collective, perez2023simulating} similar to the Holstein-Tavis-Cummings method.\cite{zeb2018exact, herrera2017dark, cui2022collective}

Following a collective perspective, we here propose a collective quantum electrodynamics coupled cluster method (C-QED-CC) that merges the \textit{ab initio} QED coupled cluster wave function (QED-CC)\cite{haugland2020coupled} with the Hopfield description of polaritons.\cite{hopfield1958theory}
The coupled cluster wave function ensures reliable modeling of electron correlation, necessary for describing realistic molecular systems, and a nonperturbative description of the electron-photon coupling (electron-photon correlation).\cite{helgaker2013molecular}
At the same time, rather than including explicitly several molecules in the simulation, we focus on one molecule treated at the \textit{ab initio} level, while the other molecules are approximated with a harmonic model.
In this way, we maintain the high wave function
accuracy within a small core that interacts with a somewhat blurred collective optical environment.
The QED-CC and Hopfield parametrizations do not rely on the rotating wave approximation and can account for dipole self-energy effects, such that the model is suited for the strong and ultrastrong coupling regimes.
The collective description of the system allows for the use of realistic values of the microscopic light-matter coupling, while the Rabi splitting can be adjusted by increasing the number of harmonic oscillators without increasing the computational costs.

Since the interplay between intermolecular forces and collective strong coupling has been claimed to produce nontrivial effects,\cite{biswas2024electronic, sidler2020polaritonic, castagnola2024collective, patrahau2024direct} we employ our model to study the impact of light-matter strong coupling on excimers.
Excimers are aggregates of two molecules that interact weakly in the ground state but form a stronger bond in the excited state.
As a case study, we focus on the argon dimer, which permits a simple analysis of the potential energy surfaces (PESs) and their vibrational levels.
Our simulations suggest that the excimer formation is inhibited in the strong coupling regime.
While the single-molecule strong coupling, modeled using QED-CC,\cite{haugland2020coupled} suggests a weakening of the excimer bond itself, the mechanism is different under collective strong coupling.
The C-QED-CC excimer PES shows characteristics from both the ground and the excited electronic states, which introduces an abrupt change in its vibrational landscape.
We argue that such a sharp change is a general feature that hinders excimer formation and points to an unexplored path for understanding chemical modifications under collective strong coupling.

The paper is organized as follows.
In \autoref{sec: theory}, we examine the Hamiltonian for a single molecule under collective strong coupling and derive the Hartree-Fock and coupled cluster models.
In \autoref{sec: results}, after validating the proposed method, we show the results for the $\text{Ar}$ dimer. 
Finally, in \autoref{sec: conclusion}, we summarize and discuss our findings.

\section{Theory} \label{sec: theory}

In this section, we derive the Hamiltonian for a molecule immersed in a collective polaritonic environment.
We show that our Hamiltonian has several physical interpretations and discuss how to include molecular disorder.
We then develop the Hartree-Fock and coupled cluster models, which are applied to the study of excimers in \autoref{sec: results}.

\subsection{Molecular Hamiltonian for collective strong coupling}

We explicitly include the electrons and nuclei of a molecule, treated within \textit{ab initio} theory, along with a harmonic oscillator $O$ of frequency $\omega_o$ and fictitious charge $Q$ and mass $M$.
The molecule and the oscillator interact with a single effective electromagnetic mode.
In Coulomb gauge and following the minimal coupling prescription, the Hamiltonian in the Born-Oppenheimer approximation reads\cite{ruggenthaler2023understanding}
\begin{align}
    &H = \frac{1}{2}\sum_i {\bigg(\bm{p}_i + \frac{1}{c}\bm{A}(\bm{r}_i)\bigg)^2} \nonumber \\
    &+\frac{1}{2}\sum_{i\neq j}\frac{1}{|\bm r_i - \bm r_j|} -\sum_{i,M}\frac{Z_M}{|\bm{r}_i-\bm{R}_M|}+\frac{1}{2}\sum_{M\neq N}\frac{Z_MZ_N}{|\bm{R}_N-\bm{R}_M|}  \nonumber \\
    & +\frac{1}{2M}{\bigg(\bm{p}_o - \frac{Q}{c}\bm{A}(\bm q_o)\bigg)^2}+\frac{M\omega_o^2}{2}\bm q_o^2 + \omega b^\dagger b,
\end{align}
where the electromagnetic field is described by a single effective mode of frequency $\omega$ with creation (annihilation) operator $b^\dagger$ ($b$), $\bm A(\bm r)$ is the vector potential computed at position $\bm r$, $c$ is the speed of light, $\bm q_o$ and $\bm p_o$ are the generalized coordinate and momenta for the oscillator $O$, $i$ and $j$ label the electrons with position and momentum $\bm r_i$ and $\bm p_i$, and $M$ and $N$ label the nuclei with charge $Z_N$ at fixed position $\bm R_N$.
The vector potential reads
\begin{equation}
{\bm{A}}(\bm r) = \frac{c\lambda}{\sqrt{2\omega}}\boldsymbol{\epsilon}\left( e^{i\bm{k}\cdot\bm{r}}b + e^{-i\bm{k}\cdot\bm{r}}b^{\dagger} \right),    
\end{equation}
where $\boldsymbol{\epsilon}$ is the polarization of the field, assumed to be real, $\bm k$ is the wave vector, and $\lambda$ is the light-matter coupling strength, connected to the effective volume confinement $V_{eff}$ of the electromagnetic field $\lambda=\sqrt{\frac{4\pi}{V_{eff}}}$.
We then resort to the long-wavelength approximation\cite{svendsen2023theory}
\begin{equation}
\bm k \cdot \bm r_i \approx \bm k \cdot \bm q_o\approx 0,
\end{equation}
so that the Hamiltonian is approximated as
\begin{align}
    &H \approx \frac{1}{2} \sum_i {\bigg(\bm{p}_i + \frac{1}{c}\bm{A}(\bm 0)\bigg)^2} \nonumber \\
    &+\frac{1}{2}\sum_{i\neq j}\frac{1}{|\bm r_i - \bm r_j|} -\sum_{i,M}\frac{Z_M}{|\bm{r}_i-\bm{R}_M|}+\frac{1}{2}\sum_{M\neq N}\frac{Z_MZ_N}{|\bm{R}_N-\bm{R}_M|}  \nonumber \\
    & +\frac{1}{2M}{\bigg(\bm{p}_o - \frac{Q}{c}\bm{A}(\bm 0)\bigg)^2}+\frac{M\omega_o^2}{2}\bm q_o^2 + \omega b^\dagger b, \label{eq: velocity HAM}
\end{align}
which is the dipole Hamiltonian in the velocity form.
We transform \autoref{eq: velocity HAM} as $e^{is}He^{-is}$, with
\begin{equation}
    s =  -   \bigg(\bm q_o Q + \bm d \bigg) \cdot \frac{\bm A(\bm 0)}{c}, \label{eq: PZW}
\end{equation}
where $\bm d$ is the dipole operator $\bm d=-\sum_i \bm r_i+\sum_N Z_N \bm R_N$. 
The transformed Hamiltonian reads
\begin{align}
    &e^{is}He^{-is} = \nonumber\\
    &\sum_i \frac{\bm{p}_i^2}{2}  +\frac{1}{2}\sum_{i\neq j}\frac{1}{|\bm r_i - \bm r_j|} -\sum_{i,M}\frac{Z_M}{|\bm{r}_i-\bm{R}_M|}+h_{nuc}  \nonumber \\
    & +\frac{\bm{p}_o^2}{2M}+\frac{M\omega_o^2}{2}\bm q_o^2+ \omega b^\dagger b  \nonumber \\
    &-i \sqrt{\frac{\omega}{2}}(b^\dagger - b) \bm \lambda \cdot \bigg(\bm q_o Q + \bm d \bigg) + \frac{1}{2} \bigg(\bm\lambda \cdot (\bm q_o Q + \bm d )\bigg)^2 ,\label{eq: Ham intermediate}
\end{align}
where $h_{nuc}$ includes the constant nuclear repulsion energy, and we defined $\bm\lambda = \lambda \boldsymbol{\epsilon}$.
We apply a phase rotation $b^\dagger \to i b^\dagger$ and define the creation and annihilation operators for the oscillator along the field polarization $ \boldsymbol{\epsilon}$ 
\begin{align}
    A &= \sqrt{\frac{M\omega_o}{2}}(\bm q_o + \frac{i}{M\omega_o}\bm p_o) \cdot \boldsymbol{\epsilon} \\
    A^\dagger &= \sqrt{\frac{M\omega_o}{2}}(\bm q_o - \frac{i}{M\omega_o}\bm p_o) \cdot \boldsymbol{\epsilon},
\end{align}
with commutation relation $[A,A^\dagger]=1$.
We thus obtain
\begin{align}
    &H = \sum_i \frac{\bm{p}_i^2}{2}  +\frac{1}{2}\sum_{i\neq j}\frac{1}{|\bm r_i - \bm r_j|} -\sum_{i,M}\frac{Z_M}{|\bm{r}_i-\bm{R}_M|}+h_{nuc}  \nonumber \\
    & +\omega_o A^\dagger A + \omega b^\dagger b \nonumber \\
    &+ g (b^\dagger + b) (A+A^\dagger) + \sqrt{\frac{\omega}{2}} (\bm{\lambda}\cdot \bm d ) (b+b^\dagger)\nonumber \\
    &+ \frac{1}{2} (\bm\lambda \cdot\bm d )^2 + \frac{g^2}{\omega}(A+A^\dagger)^2+ g \sqrt{\frac{2}{\omega}} (\bm{\lambda}\cdot \bm d ) (A+A^\dagger) ,\label{eq: Ham final}
\end{align}
where we have defined an effective coupling $g$ between the photons and the oscillator
\begin{equation}
    g = \lambda \sqrt{\frac{\omega}{2}} \frac{ Q}{\sqrt{2\omega_oM}} = \frac{1}{2}\sqrt{\frac{\omega}{\omega_o}} \frac{\lambda Q}{\sqrt{M}}.
\end{equation}
In \autoref{eq: Ham final}, the first line is the standard electronic Hamiltonian of a molecule in the Born-Oppenheimer approximation, the second line includes the energies of the electromagnetic field and the oscillator $O$, the third line includes the coupling of $O$ and the molecule with the photon field, and the last line is the dipole self-energy of the system.
In the last line of \autoref{eq: Ham final}, a coupling term between $O$ and the molecule appears
\begin{equation}
    g \sqrt{\frac{2}{\omega}} (\bm{\lambda}\cdot \bm d ) (A+A^\dagger).\label{eq: apparent coupling} 
\end{equation}
The apparent coupling in \autoref{eq: apparent coupling} is a consequence of the picture change via \autoref{eq: PZW}.
In fact, \autoref{eq: Ham intermediate} shows that it originates from the dipole self-energy.
Retaining this dipole self-energy contribution is necessary to ensure manifestly origin-invariant energies for charged systems.
The oscillator dipole self-energy can be reabsorbed by a Bogoliobov transformation
\begin{equation}
U = \textrm{exp}\left[\frac{\chi}{2}\left(\big(A^{\dagger}\big)^{2}-A^2\right)\right],   
\label{eq:Rotation}
\end{equation}
with $\tanh 2\chi = \frac{2 g^2}{2g^2+ \omega \omega_o}$.
The resulting Hamiltonian reads
\begin{align}
    H& = \sum_i \frac{\bm{p}_i^2}{2}  +\frac{1}{2}\sum_{i\neq j}\frac{1}{|\bm r_i - \bm r_j|} -\sum_{i,M}\frac{Z_M}{|\bm{r}_i-\bm{R}_M|}+h_{nuc}  \nonumber \\
    & +\tilde \omega_o A^\dagger A + \omega b^\dagger b \nonumber \\
    &+ \tilde g (b^\dagger + b) (A+A^\dagger) + \sqrt{\frac{\omega}{2}} (\bm{\lambda}\cdot \bm d ) (b+b^\dagger)\nonumber \\
    &+ \frac{1}{2} (\bm\lambda \cdot\bm d )^2+ \tilde g \sqrt{\frac{2}{\omega}} (\bm{\lambda}\cdot \bm d ) (A+A^\dagger), \label{eq: Ham final reabs dse}
\end{align}
where the oscillator frequency and light-matter coupling have been renormalized
\begin{align}
     \tilde \omega_o &= \sqrt{\omega_o^2 + 4 g^2 \frac{\omega_o}{\omega}},\qquad
    \tilde g = g \sqrt{\frac{\omega_o}{\tilde \omega_o}}.
\end{align}

The Hamiltonian in \autoref{eq: Ham final reabs dse} is derived for a molecule in an optical cavity by including a single photon mode and an effective harmonic oscillator.
An analogous Hamiltonian can be obtained by considering explicitly $N$  harmonic toy molecules, which constitute the collective polaritonic environment.
This Hamiltonian, in second quantization and length form, is given by
\begin{align}
    H&=\sum_{pq}h_{pq}E_{pq}+\frac{1}{2}\sum_{pqrs}g_{pqrs}e_{pqrs}+\omega{b}^\dagger{b}+\mathcal{E}\sum_{n=1}^{N} a^\dagger_n a_n\nonumber\\
    &+\sqrt{\frac{\omega}{2}}\sum_{pq}(\bm{\lambda}\cdot \bm{d})_{pq}E_{pq}(b^\dagger+b)+
    \sum_{n=1}^{N} g_n (a_n+a^\dagger_n)(b+b^\dagger) \nonumber\\
    &
    +\frac{1}{\omega}\bigg[\sum_{n=1}^{N}g_n(a_n+a^\dagger_n)\bigg]^2\nonumber
    +\frac{1}{2}\sum_{pqrs}(\bm{\lambda}\cdot \bm{d})_{pq}(\bm{\lambda}\cdot \bm{d})_{rs}E_{pq}E_{rs}\nonumber\\
    &+\sqrt{\frac{2}{\omega}}\sum_{n=1}^N g_n (a_n+a^\dagger_n) \sum_{pq}(\bm{\lambda}\cdot \bm{d})_{pq}E_{pq}. \label{eq: Hop env}
\end{align}
Here,  $E_{pq}$ and $e_{pqrs}$ are the spin-adapted one and two-electron singlet operators for the \textit{ab initio} molecule in an orthonormal orbital basis indexed by $p,q,r,s$.\cite{helgaker2013molecular}
All the toy molecules are assigned creation (annihilation) operators $a_n^\dagger$ ($a_n$), harmonic energy $\mathcal{E}$, and light-matter coupling strength $g_n$.
We then define the collective coupling $\bar g$\cite{houdre1996vacuum}
\begin{equation}
    \bar g = \sqrt{\sum_{n=1}^N g_n^2}
\end{equation}
and introduce the collective operator
\begin{equation}
    \bar A = \frac{1}{\bar g}\sum_{n=1}^N {g_n} \,a_n, \label{eq: collective right mode}
\end{equation}
together with $(N-1)$ dark state operators $D_k=\sum_{n=1}^N X_{kn} a_n$ for $k = 1,\,\dots,\, (N-1)$,
where the rows $\vec R_k^T = (X_{kn})$ form an orthonormal set chosen to be perpendicular to the vector $\vec R_0 = ({g_n})$.
Using this change of basis, the Hamiltonian can be written as
\begin{align}
    H&=\sum_{pq}h_{pq}E_{pq}+\frac{1}{2}\sum_{pqrs}g_{pqrs}e_{pqrs}\nonumber\\
    &+\frac{1}{2}\sum_{pqrs}(\bm{\lambda}\cdot \bm{d})_{pq}(\bm{\lambda}\cdot \bm{d})_{rs}E_{pq}E_{rs}\nonumber\\
    &+\omega{b}^\dagger{b}+\mathcal{E}\sum_{k=1}^{N-1} D^\dagger_k D_k + \mathcal{E} \bar A^\dagger \bar A\nonumber\\
    &+\sqrt{\frac{\omega}{2}}\sum_{pq}(\bm{\lambda}\cdot \bm{d})_{pq}E_{pq}(b^\dagger+b)+
    \bar g (\bar A+\bar A^\dagger)(b+b^\dagger) \nonumber\\
    &
    +\frac{\bar g^2}{\omega}(\bar A+\bar A^\dagger)^2+\sqrt{\frac{2}{\omega}}\bar g (\bar A+\bar A^\dagger) \sum_{pq}(\bm{\lambda}\cdot \bm{d})_{pq}E_{pq}. \label{eq: Hop env many}
\end{align}
With the identification
\begin{align}
     \bar g &\longleftrightarrow g ,\qquad
     \bar A \longleftrightarrow A ,\qquad
     \mathcal{E} \longleftrightarrow \omega_o
\end{align}
and ignoring the dark state operators $D_k$, which do not couple to the photon or the molecule, we see that \autoref{eq: Hop env many} and \autoref{eq: Ham final} describe the same system.
We then interpret the oscillator $O$ in \autoref{eq: Ham final} as the collective response of all the remaining molecules in the optical cavity.
Therefore, while $\lambda$ is a microscopic (single-molecule) parameter, $g$ is a collective (macroscopic) parameter $g\sim \lambda \sqrt{N} \gg \lambda$.
Notice also that, since in \autoref{eq: Hop env many} there is only one relevant mode coupled to the field (that is, the bright collective mode in \autoref{eq: collective right mode}), increasing $N$ does not correspond to any extra computational demands.
Since the harmonic molecules are assumed to have the same excitation energy $\mathcal{E}$, our model disregards inhomogeneous broadening.
Nevertheless, we take positional and orientational disorder into account by using a different coupling strength $g_n$ for each toy molecule. 
The energetic disorder has been suggested to have a relevant impact on polaritonic properties,\cite{dutta2024thermal} and we could introduce it by including explicitly several independent replicas of the molecular system.
This procedure has a steep computational cost, defined by the \textit{ab initio} molecule.
A less refined modeling of the energetic disorder within the presented framework can be achieved by using an extended set of oscillators $\{O_s\}_{s=1}^S$ in \autoref{eq: velocity HAM}, assigning to each a different excitation energy $\omega_o^s$ close to $\omega_o$. \\

Finally, following the discussion of Hopfield on polaritons,\cite{hopfield1958theory} the Hamiltonian in \autoref{eq: Ham final} can be derived from the Lagrangian of a molecule and a classical dielectric, interacting with the electromagnetic field.
The dielectric is described by a polarization density $\bm{P}$ with a linear restoring force
\begin{equation}
    \frac{\partial^2\bm P}{\partial t^2}+\omega_o^2\bm P = \omega_o^2\beta  \bm E ,\label{eq: constit relation dielectric}
\end{equation}
where $\bm E$ is the electric field.
This equation determines a dielectric dispersion law as in the Drude model\cite{bade1957drude, hopfield1958theory}
\begin{equation}
    \epsilon(\omega) = 1 + \frac{4\pi \beta}{1- \omega^2/\omega_o^2}.
\end{equation}
We can thus think of \autoref{eq: Ham final} as a molecule in a dielectric environment dressed by the electromagnetic field.
A more detailed derivation is presented in Sec. S1 of the Supporting Infomation. \\

The Hamiltonian we employ can thus be interpreted as 1) describing a molecule and an effective oscillator coupled to the electromagnetic field (\autoref{eq: Ham final reabs dse}), 2) describing explicitly one \textit{ab initio} molecule with $N$ toy-model molecules coupled to the field (\autoref{eq: Hop env many}), or 3) describing a molecule and a classical Drude dielectric in an optical cavity.
A similar Hamiltonian can be obtained starting from the Tavis-Cummings model,\cite{dicke1954coherence, tavis1968exact} where the $N$ environment molecules are treated as two-level systems instead of harmonic oscillators as in \autoref{eq: Hop env} (see e.g. Ref.\cite{sokolovskii2024one}).
However, using a harmonic model, we do not need to resort to the rotating wave approximation, and we can include some of the dipole self-energy effects arising from the environment molecules.
The transformation in \autoref{eq: PZW} also introduces some symmetry between the oscillator $O$ and the photon field, so implementing equations for the Hamiltonian in \autoref{eq: Ham final reabs dse} is very similar to a multi-mode \textit{ab initio} QED theory.
Moreover, although we here focus on electronic strong coupling, a harmonic description is a natural choice also for vibrational strong coupling.
Our approach thus presents several advantages for an \textit{ab initio} treatment in the collective strong coupling regime.


\subsection{Hartree-Fock}

We now discuss a mean-field approximation, based on the Hamiltonian in \autoref{eq: Ham final reabs dse}, which serves as a reference for coupled cluster theory.
The reference wave function $\ket{\text R}$ is a tensor product state
\begin{equation}
    \ket{\text R} = \ket{\text S} \otimes \ket{P}\otimes \ket{E},
\end{equation}
where $\ket{\text S}$ is a Slater determinant for the molecule and $\ket{P}$ is a photon state expressed in terms of expansion coefficients $c_n$
\begin{equation}
    \ket{P} = \sum_n \big(b^\dagger\big)^n\ket{0}c_n ,
\end{equation}
where $\ket{0}$ is the electromagnetic vacuum.
Finally, $\ket{E}$ is an oscillator state with expansion coefficients $d_n$
\begin{equation}
    \ket{E} = \sum_n \big(A^\dagger\big)^n\ket{0}d_n ,
\end{equation}
where $\ket{0}$ is the oscillator ground state.
The optimal parameters are obtained by minimizing the expectation value of the Hamiltonian with respect to $c_n$, $d_n$, and the orbitals in the Slater determinant.
The optimal state is given by the QED-HF Slater determinant and its corresponding coherent state for the photon\cite{haugland2020coupled} in the environment ground state (the full derivation is presented in the Supporting Information)
\begin{align}
    \ket{\text R}& = \ket{\text{HF}} \otimes \text{exp}\bigg( - \frac{\bm \lambda \cdot \braket{\bm d}_\text{HF}}{\sqrt{2\omega}}(b^\dagger - b)\bigg) \ket{0} \otimes \ket{0}\\
    &\equiv U_{\text{QED-HF}}\ket{\text{HF},0,0}.
\end{align}
Transforming the Hamiltonian in \autoref{eq: Ham final reabs dse} with $U_{\text{QED-HF}}$ we get
\begin{align}
    &U^\dagger_{\text{QED-HF}}H\,U_{\text{QED-HF}} = \nonumber\\
    &\sum_i \frac{\bm{p}_i^2}{2}  +\frac{1}{2}\sum_{i\neq j}\frac{1}{|\bm r_i - \bm r_j|} -\sum_{i,M}\frac{Z_M}{|\bm{r}_i-\bm{R}_M|}+h_{nuc}  \nonumber \\
    & + \frac{1}{2} (\bm\lambda \cdot(\bm{d}-\braket{\bm{d}}_{\text{QED-HF}}) )^2+\tilde \omega_o A^\dagger A + \omega b^\dagger b \nonumber \\
    &+ \tilde g (b^\dagger + b) (A+A^\dagger) + \sqrt{\frac{\omega}{2}} (\bm{\lambda}\cdot (\bm{d}-\braket{\bm{d}}_{\text{QED-HF}}) ) (b+b^\dagger)\nonumber \\
    &+ \tilde g \sqrt{\frac{2}{\omega}} (\bm{\lambda}\cdot (\bm{d}-\braket{\bm{d}}_{\text{QED-HF}}) ) (A+A^\dagger) , \label{eq: Ham coherenst transf}
\end{align}
which is manifestly origin invariant even for charged systems.

\subsection{Coupled cluster}

The coupled cluster wave function $\ket{\text{CC}}$, in the coherent-state representation defined in \autoref{eq: Ham coherenst transf}\cite{helgaker2013molecular, haugland2020coupled}, is given by
\begin{equation}
    \ket{\text{CC}} = \text{exp}(T)\ket{\text{HF},0,0},
\end{equation}
where $T$ is the cluster operator, including excitations of the molecule, the photon field, and the oscillator $O$.
In the following, we outline what we refer to as C-QED-CCSD-1-1 theory. 
The cluster operator $T$ includes single and double excitations of the molecule and a single excitation of the oscillator and the photon field
\begin{align}
    T&=\sum_{ai}t_{ai}E_{ai}+\frac{1}{2}\sum_{aibj}t_{aibj}E_{ai}E_{bj} \nonumber\\
    &+\sum_{ai}s_{ai}E_{ai}b^\dagger + 
    \frac{1}{2}\sum_{aibj}s_{aibj}E_{ai}E_{bj}b^\dagger\nonumber\\
    &+ \gamma b^\dagger + \theta A^\dagger + \kappa A^\dagger b^\dagger \nonumber \\
    &+\sum_{ai}o_{ai}E_{ai}A^\dagger + 
    \frac{1}{2}\sum_{aibj}o_{aibj}E_{ai}E_{bj}A^\dagger \nonumber \\
    &+\sum_{ai}\rho_{ai}E_{ai}A^\dagger b^\dagger + 
    \frac{1}{2}\sum_{aibj}\rho_{aibj}E_{ai}E_{bj}A^\dagger b^\dagger , \label{eq: cluster T}
\end{align}
where $i$ and $j$ ($a$ and $b$) label occupied (virtual) orbitals relative to the reference Slater determinant.
The cluster operator thus includes the electronic single and double excitation amplitudes $t_{ai}$ and $t_{aibj}$, the photon, oscillator, and oscillator-photon excitation amplitudes $\gamma$, $\theta$, and $\kappa$, the electron-photon excitation amplitudes $s_{ai}$ and $s_{aibj}$, the electron-oscillator excitation amplitudes $o_{ai}$ and $o_{aibj}$, as well as the amplitudes $\rho_{ai}$ and $\rho_{aibj}$ describing the simultaneous excitation of the molecule, the oscillator, and the photon field.
We can simplify the operator $T$ working on the electron-oscillator amplitudes.
For instance, retaining only the first three lines of \autoref{eq: cluster T}, we obtain a minimal cluster operator describing the oscillator coupled to the photon field while we disregard the simultaneous excitations of the molecule and the oscillator, i.e., we disregard electron-oscillator correlation.
If we keep the first four lines, we reintroduce electron-oscillator correlation but disregard the simultaneous excitation of all three components of our system. 
Nevertheless, in this study, we keep the complete C-QED-CCSD-1-1 cluster operator in \autoref{eq: cluster T}, and the validity of truncating $T$ will be studied elsewhere.

The ground state energy and amplitudes are obtained by projection of the Schr\"odinger equation onto the states $\ket{\text R}$ and $\tau_\mu\ket{\text R}=\ket{\mu}$\cite{helgaker2013molecular, haugland2020coupled}
\begin{align}
    &E_\text{CC}=\braket{\text R |H \text{exp}(T)|\text R}\\
    &\Omega_{\mu}=\braket{\mu |\text{exp}(-T)H \text{exp}(T)|\text R}=0,
\end{align}
 where $\tau_\mu$ is an excitation operator included in $T=\sum_\mu t_\mu \tau_\mu$ .
The excited state properties are obtained within the equation-of-motion (EOM) formalism, where the excitation energies and the excited states are determined by diagonalizing the coupled cluster Jacobian $\textbf{A}$
\begin{equation}
    \text A_{\mu,\nu}=\braket{\mu|\text{exp}(-T) [H, \tau_\nu] \text{exp}(T)|\text R} , \label{eq: jacobian}
\end{equation}
as in electronic coupled cluster theory.\cite{helgaker2013molecular, haugland2020coupled}

The proposed coupled cluster method thus allows us to obtain the properties of a molecule immersed in a structured electromagnetic environment in the collective strong coupling regime, including electron-electron, electron-photon, and electron-collective-environment correlation, providing a reliable \textit{ab initio} simulation.
In the next section, we validate the proposed model and analyze the effect of collective strong coupling on the properties of excimers.

\section{Results}\label{sec: results}
The C-QED-CCSD-1-1 ground state and EOM equations are implemented in a local branch of the $e^\mathcal{T}$ program, an open-source electronic structure program.\cite{folkestad2020t}
First, we validate the proposed wave function method by comparing it to the results presented in Ref.\cite{castagnola2024collective}
Then, we apply the developed method to the properties of excimers, studying $\text{Ar}_2$.

\subsection{Validation of the theory}
\begin{figure*}
    \centering
    \includegraphics[width=\textwidth]{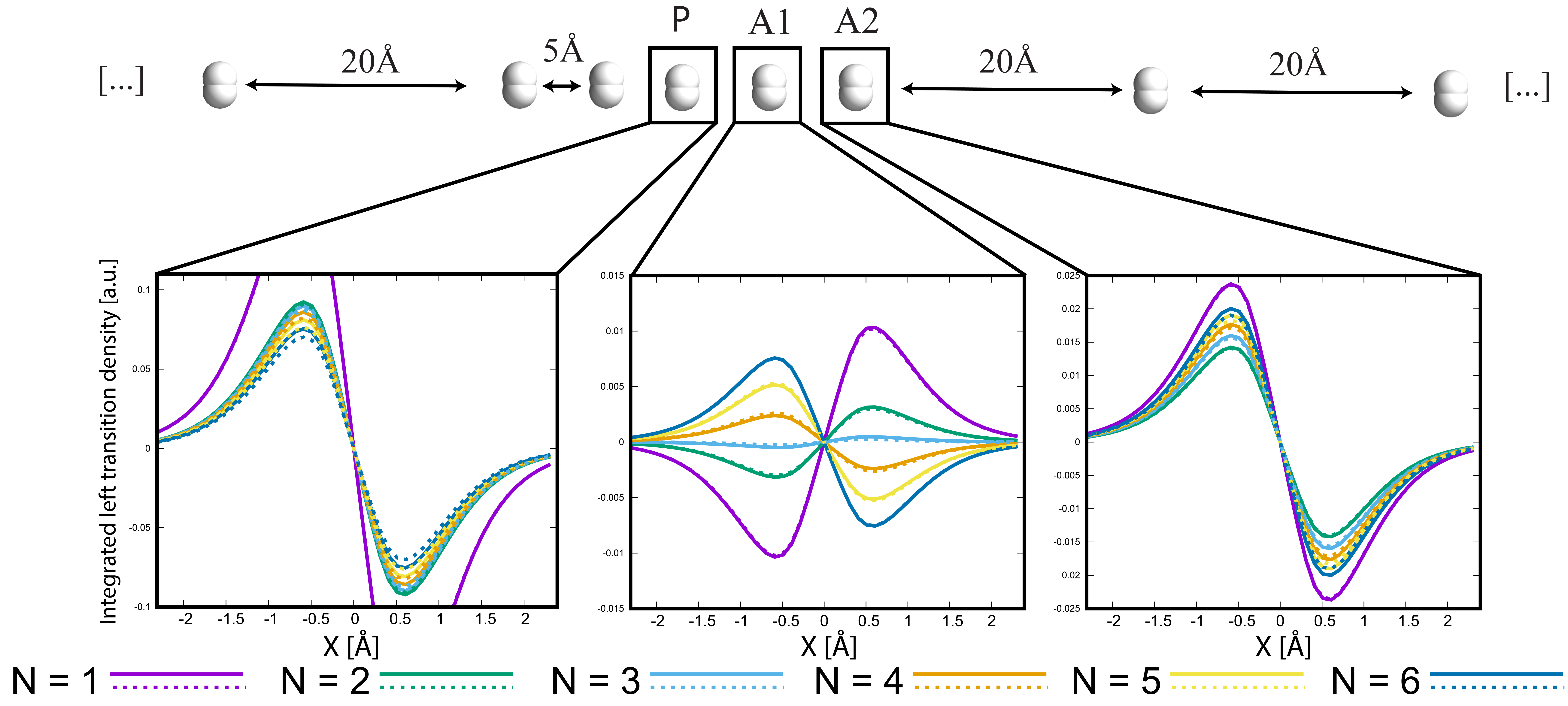}
    \caption{Pictorial representation of the hydrogen cluster ($\text H_2)_5$ in the presence of $N$ additional farther away $\text H_2$ molecules which couple to the optical device forming what we refer to as the collective polaritonic environment. 
    The central molecule, labeled P, is slightly stretched compared to the others and provides a model for an impurity or a nucleation center.\cite{castagnola2024collective} 
    The two adjacent molecules, labeled A1 and A2, are placed at \qty[mode = text]{5}{\angstrom} distance, and thus intermolecular forces are significant. 
    The distant $N$ molecules are placed at \qty[mode = text]{20}{\angstrom} from  $(\text H_2)_5$ and each other, and directly interact only with the photon field. The photon is set to be resonant with the bare $\text H_2$ excitation, with a light-matter coupling strength of $\lambda = $ \qty[mode = text]{0.005}{\atomicunit} (additional computational details can be found in Sec. S2 in the Supporting Information). The local properties of P, A1, and A2 are obtained from the left transition density of the lower polariton (LP), integrated around each molecule in the directions perpendicular to the bond. The QED-CCSD results, in which the far away molecules are included explicitly, are presented as reference (dotted lines), while solid lines are used for the C-QED-CCSD-1-1 model based on \autoref{eq: Hop env many} with $N=1-6$. The two models give very similar results, predicting an inversion in the LP local transition density of A1 as a function of the collective strong coupling. Notice that while the computational demands (both in terms of memory and CPU time) for the QED-CCSD calculations increase steeply when adding more molecules to the simulation, the complexity of the C-QED-CCSD-1-1 method is independent of $N$.}
    \label{fig: H2 paper}
\end{figure*}
The wave function we propose aims at reproducing the behavior of a molecule in an optical device, where several other molecules are simultaneously (collectively) coupled to the electromagnetic modes of the resonator.
To validate our method, we first focus on the results obtained in Ref.\cite{castagnola2024collective}, where the local transition properties of a hydrogen cluster $(\text H_2)_5$ are studied by adding additional $\text H_2$ molecules far away.
The molecules in the cluster interact via intermolecular forces, and the distant $\text H_2$ molecules contribute only through their collective coupling to the field.
The transition densities computed in Ref.\cite{castagnola2024collective} have been recalculated using QED-CCSD and are shown in \autoref{fig: H2 paper} (dotted lines) for $\lambda = $ \qty[mode = text]{0.005}{\atomicunit} and intermolecular distance of \qty[mode = text]{5}{\angstrom} in the central $(\text H_2)_5$ cluster.
The central molecule in $(\text H_2)_5$, labeled P, has a slightly stretched bond length, and the other $N$ $\text H_2$ molecules are added at \qty[mode = text]{20}{\angstrom} distance from each other and the cluster, as shown pictorially in \autoref{fig: H2 paper} (additional computational details are provided in Sec. S2 of the Supporting Information).
The results show that the lower polariton (LP) local transition density of the $\text H_2$ labeled A1, adjacent to the stretched central dimer P, changes sign with increasing $N$. 
That is, the local response property in the LP of the first polarization shell of an impurity is qualitatively modified by collective strong coupling.\cite{castagnola2024collective}
Using our C-QED-CCSD-1-1 model, similar calculations are performed retaining the central $(\text H_2)_5$ cluster and modeling the distant $\text H_2$ molecules as oscillators using \autoref{eq: Hop env many} for $N=1-6$.
The oscillator energy and the coupling strength $g$ are parametrized from an electronic CCSD calculation on $\text H_2$.
The C-QED-CCSD-1-1 results in \autoref{fig: H2 paper} (solid lines) show good agreement with the QED-CCSD calculations (dotted lines) in which the additional $\text H_2$ molecules are treated explicitly.
The minor differences are likely a consequence of the harmonic and single-state description of the distant $\text H_2$ molecules, contrary to the QED-CCSD calculation where the whole excited state manifolds are implicitly included in the full \textit{ab initio} description. 
In \autoref{tab: polariton energies}, we report the excitation energies for the lower, middle, and upper polaritons computed using QED-CCSD and C-QED-CCSD-1-1.
\begin{table}[!ht]
    \centering
\setlength{\tabcolsep}{8pt} 
\renewcommand{\arraystretch}{1.3} 
    \caption{Lower, middle, and upper polariton excitation energies of the $(\text H_2)_5$ cluster in the collective environment shown in \autoref{fig: H2 paper} computed with the QED-CCSD (retaining explicitly the far away $\text H_2$ molecules) and C-QED-CCSD-1-1 (using fictitious harmonic oscillators) models.
    The table shows that there is a good quantitative agreement between the simulated energies.
    }
    \begin{tabular}{ccc}
    \hline
 \multicolumn{3}{ c }{Lower polariton -- LP}\\
        \hline  N  & QED-CCSD [eV]  &  C-QED-CCSD-1-1 [eV] \\
       \hline 1 &12.2948&12.2949\\
       2 &12.2913&12.2915\\
      3  &12.2869&12.2873\\
      4  &12.2814&12.2821\\
      5  &12.2751&12.2761\\
      6  &12.2681&12.2694\\ 
    \hline
 \multicolumn{3}{ c }{Middle polariton -- MP}\\
        \hline N  & QED-CCSD [eV]  &  C-QED-CCSD-1-1 [eV] \\
        \hline 1 &12.3694&12.3695\\
      2  &12.3573&12.3577\\
      3  &12.3478&12.3483\\
      4  &12.3404&12.3409\\
      5  &12.3348&12.3353\\
      6  &12.3306&12.3311\\
    \hline
 \multicolumn{3}{ c }{Upper polariton -- UP}\\
        \hline N  & QED-CCSD [eV]  &  C-QED-CCSD-1-1 [eV] \\
        \hline 1 &12.6680&12.6682\\
       2 &12.6792&12.6797\\
       3 &12.6898&12.6907\\
      4  &12.6999&12.7012\\
      5  &12.7096&12.7113\\
      6  &12.7189&12.7210\\
        \hline
    \end{tabular}
    \label{tab: polariton energies}
\end{table}
The good agreement between the simulations, both for global excitation energies and local transition properties, validates our coupled cluster wave function for collective strong coupling.
Additional results are presented in Sec. S3 of the Supporting Information, demonstrating that we replicate all the effects outlined in Ref.\cite{castagnola2024collective}, and furthermore expand the discussion for large $N$ and small $\lambda$.

Having validated the proposed method, we now discuss how collective strong coupling modifies the properties of excimers.

\subsection{Argon excimer}

\begin{figure}[!ht]
    \centering
    \includegraphics[width=.5\textwidth]{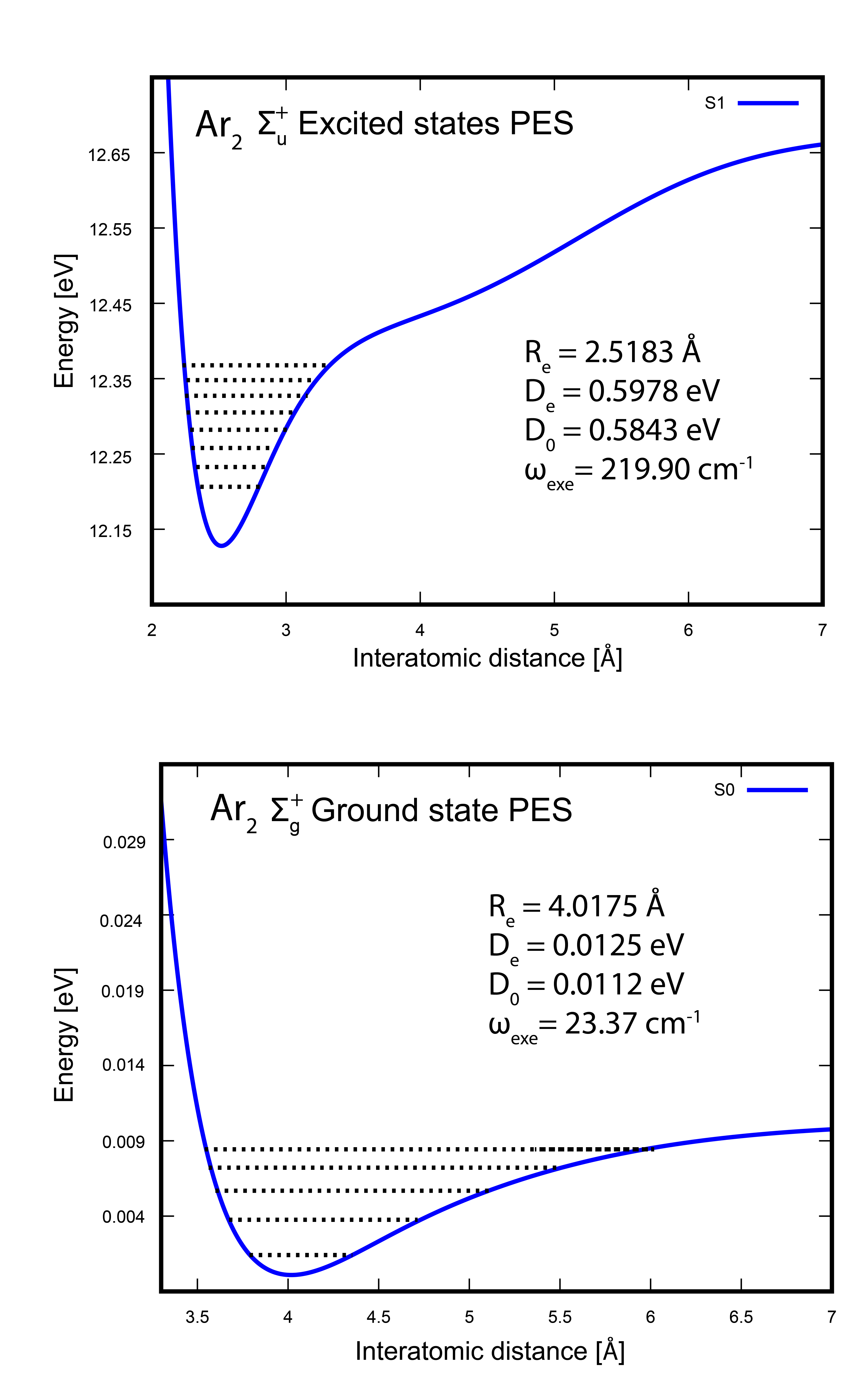}
    \caption{Upper panel: potential energy surface (PES) of the first excited singlet $\text{S}_1$ of $\text{Ar}_2$ as a function of the intermolecular separation.
    Lower panel: ground state $\text{S}_0$ PES of $\text{Ar}_2$.
    For both PESs, we report properties computed using the VIBROT module of the OpenMolcas program \cite{aquilante2020modern}.
    The dissociation energies, $D_e$ and $D_0$, and the harmonic frequency $\omega_{exe}$ show that the molecule in the ground state is only slightly bound. 
    In the excited state $\text{S}_1$, the $\text{Ar}_2$ dimer forms a more stable excimer.}
    \label{fig: Ar2 electronic}
\end{figure}

As a case study, we analyze the consequences of light-matter strong coupling on the $\text{Ar}_2$ excimer, extensively studied experimentally and theoretically.\cite{fernandez1998accurate, woon1994benchmark, mizukami1990potential, yates1983b, aziz1993highly, colbourn1976spectrum, signorell1997first, herman1988vacuum}
In \autoref{fig: Ar2 electronic}, we report the ground state $\text{S}_0$ and first excited state $\text{S}_1$ potential energy surfaces (PESs) for $\text{Ar}_2$ as a function of the intermolecular separation, calculated with CCSD using aug-cc-pVDZ basis set.
The vibrational analysis of the PES is computed using the VIBROT module of the OpenMolcas program.\cite{aquilante2020modern}
We report the first vibrational levels, the equilibrium position $R_e$, the dissociation energies ($D_e$ from the minimum of the PES, $D_0$ including the zero-point vibrational energy), and the harmonic vibrational frequency $\omega_{exe}$ of $\text{S}_0$ and $\text{S}_1$.
The ground state is only slightly bound, with an equilibrium bond length $\sim$\qty{4}{\angstrom}, harmonic frequency $\omega_{exe}$ of only \qty{23.4}{\per\centi\metre}, and dissociation energy of about \qty{0.01}{\electronvolt}.
Inspection of the $\text S_0$ vibrational wave functions reveals they are spread over considerable distances, even for the first vibrational excitations (notice also that, at room temperature $\mathcal T$, $k_B \mathcal T \sim $ \qty{25}{\milli\electronvolt} $\gg D_e\sim $ \qty{10}{\milli\electronvolt}, where $k_B$ is the Boltzmann constant).
On the other hand, the first excited state $\text{S}_1$ has a deeper potential well, with dissociation energy above \qty{0.5}{\electronvolt}, harmonic frequency $\omega_{exe}=$ \qty{219.9}{\per\centi\metre}, and an equilibrium bond length of \qty{2.51}{\angstrom}.
Compared to the experimental values for the $\text{Ar}$ dimer,\cite{aziz1993highly, colbourn1976spectrum, signorell1997first, herman1988vacuum} we slightly overestimate the bond lengths and underestimate the $\text{S}_1$ dissociation energy.
Since the PESs are dominated by dispersion forces, large basis sets and more accurate methods such as CCSD(T) or CC3 would be needed for reproducing the experimental data, and several \textit{ab initio} studies are available.\cite{fernandez1998accurate, woon1994benchmark, mizukami1990potential, yates1983b}
Spin-orbit coupling is also known to be relevant for $\text{Ar}_2$.
Nevertheless, the results are in qualitative agreement with experimental data. 

In \autoref{tab: GS AR2 parameters}, we report the computed vibrational constants for the ground state using QED-CCSD and C-QED-CCSD-1-1 for different light-matter coupling strengths $\lambda$.
The C-QED-CCSD-1-1 results are obtained by setting the microscopic light-matter coupling strength to $\lambda_0=$ \qty{0.0001}{\atomicunit} and $N$ such that $\lambda =\lambda_0\sqrt{N}\sim g$ is the same as the corresponding QED-CCSD calculation.
The harmonic energies and oscillator strengths for C-QED-CCSD-1-1 are parametrized from the electronic CCSD simulation of a single $\text{Ar}$ atom.
The photon energy is set to be resonant to the first excitation of $\text{Ar}$, and the field polarization is along the $\text{Ar}_2$ bond.
As can be seen from \autoref{tab: GS AR2 parameters}, for QED-CCSD, increasing the microscopic light-matter coupling strength $\lambda$ does lead to a (slight) modification of the ground state properties (see also Ref. \cite{haugland2021intermolecular}).
\begin{table}[!ht]
    \centering
\setlength{\tabcolsep}{8pt} 
\renewcommand{\arraystretch}{1.3} 
    \caption{Equilibrium position $R_e$, dissociation energies ($D_e$ from the PES minimum, $D_0$ including the zero-point vibrational energy), and harmonic vibrational frequency $\omega_{exe}$ computed at different light-matter coupling strengths for QED-CCSD and C-QED-CCSD-1-1.
    }
    \begin{tabular}{ccccc}
       \hline$\lambda$ [a.u.]  & $R_e$ [\qty{}{\angstrom}]&$D_e$ [eV]&$D_0$ [eV]&$\omega_{exe}$ [cm$^{-1}$]\\
    \hline
 \multicolumn{5}{ c }{QED-CCSD}\\
    \hline  0.01 &4.0193&0.0124&0.0111&23.31\\
      0.025  &4.0285&0.0119&0.0106&22.98\\
    \hline
 \multicolumn{5}{ c }{C-QED-CCSD-1-1}\\
    \hline  0.01 &4.0175&0.0127&0.0114&23.37\\
      0.025  &4.0175&0.0127&0.0114&23.37\\
      0.04  &4.0175&0.0126&0.0112&23.37\\ 
        \hline
    \end{tabular}
    \label{tab: GS AR2 parameters}
\end{table}
In particular, QED-CCSD predicts that the $\text{Ar}$ dimer ground state is destabilized by the coupling to the photon, as evidenced by a longer equilibrium bond length, reduced dissociation energies, and smaller harmonic frequency.
Nevertheless, the computed changes are very small.
From the C-QED-CCSD-1-1 calculation, for which the coupling is collective, we see that no modifications occur.
This is reasonable since from \autoref{eq: Ham final reabs dse}, the leading perturbative correction to the uncoupled electronic ground state is quadratic in the microscopic light-matter coupling strength $\lambda_0$. \\ 

\begin{figure*}
    \centering
    \includegraphics[width=\textwidth]{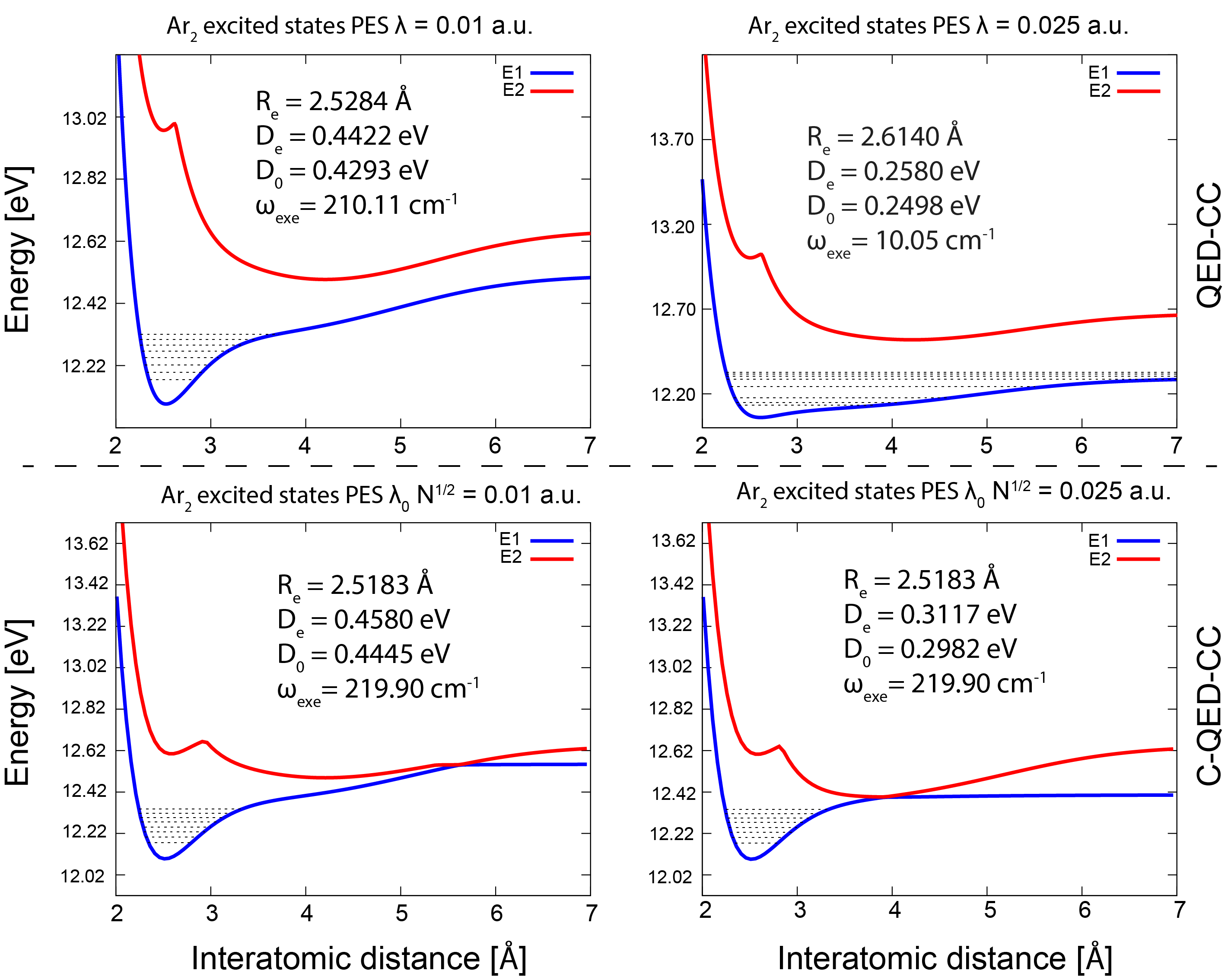}
    \caption{Potential energy surfaces of the first two excited states, E1 and E2, for $\text{Ar}_2$ using QED-CCSD (upper panels) and C-QED-CCSD-1-1 (lower panels).
    The collective coupling strength in the C-QED-CCSD-1-1 calculations is $\lambda_0 \sqrt{N}$ where $\lambda_0=$ \qtylist{0.0001}{\atomicunit} and $N$ is chosen to achieve the same couplings as in the QED-CCSD calculations ($\lambda=$ \qtylist{0.01; 0.025}{\atomicunit}).
    While the QED-CCSD calculations show a significant modification of the properties of the PESs, the C-QED-CCSD-1-1 vibrational constants are fundamentally unchanged compared to the electronic simulations. 
    In the QED-CCSD results, the first two excited states are energetically well separated, while in the C-QED-CCSD-1-1 case, they show a small avoided crossing at a distance $R_A$ defined by the collective coupling strength.
    }
    \label{fig: Ar2 excited states}
\end{figure*}

The differences are more striking for the excited states of the system.
In \autoref{fig: Ar2 excited states}, we plot the first two excited states E1 and E2, computed using QED-CCSD ($\lambda=$ \qtylist{0.01;0.025}{\atomicunit}) and C-QED-CCSD-1-1 ($\lambda_0\sqrt{N}=$ \qtylist{0.01;0.025}{\atomicunit}).
We also report the vibrational constants for the first excited state E1.
The two methods show very different results.
Using QED-CCSD and a considerable microscopic light-matter coupling strength (upper panels of \autoref{fig: Ar2 excited states}) profoundly modifies the PES compared to the electronic results in \autoref{fig: Ar2 electronic}.
The equilibrium bond length is shifted to larger distances with increasing $\lambda$, and the harmonic vibrational frequency decreases significantly.
Therefore, the PES is fundamentally changed even around the local minimum. 
The dissociation energies decrease, and the properties of all the vibrational states $\nu$ are changed accordingly (for instance, the PES becomes highly anharmonic for $\lambda=$ \qtylist{0.025}{\atomicunit}).
Overall, the light-matter coupling results in a weakening of the excimer bond.
At the same time, the second excited state E2, is energetically well separated from E1 for all the intermolecular distances $R$. 

We observe a different behavior under collective strong coupling, described using C-QED-CCSD-1-1 (bottom panels of \autoref{fig: Ar2 excited states}).
The bond length $R_e$ and the harmonic frequency $\omega_{exe}$ are the same as in the electronic structure simulation without light-matter coupling in \autoref{fig: Ar2 electronic}.
The dissociation energies are modified, but the zero-point vibrational energy $(D_0-D_e)$ is the same as \autoref{fig: Ar2 electronic}.
Therefore, we predict no modifications of the wave function around the minimum $R_e$.
In addition, we see that the first two excited states are not well separated, contrary to the QED-CCSD simulation, and show a (very small) avoided crossing at a distance $R_A$ defined by the collective coupling.
Thus, the upper and lower panels of \autoref{fig: Ar2 excited states} depict physically different situations, which would be observable under appropriate experimental conditions. \\

\begin{figure*}
    \centering
    \includegraphics[width=\textwidth]{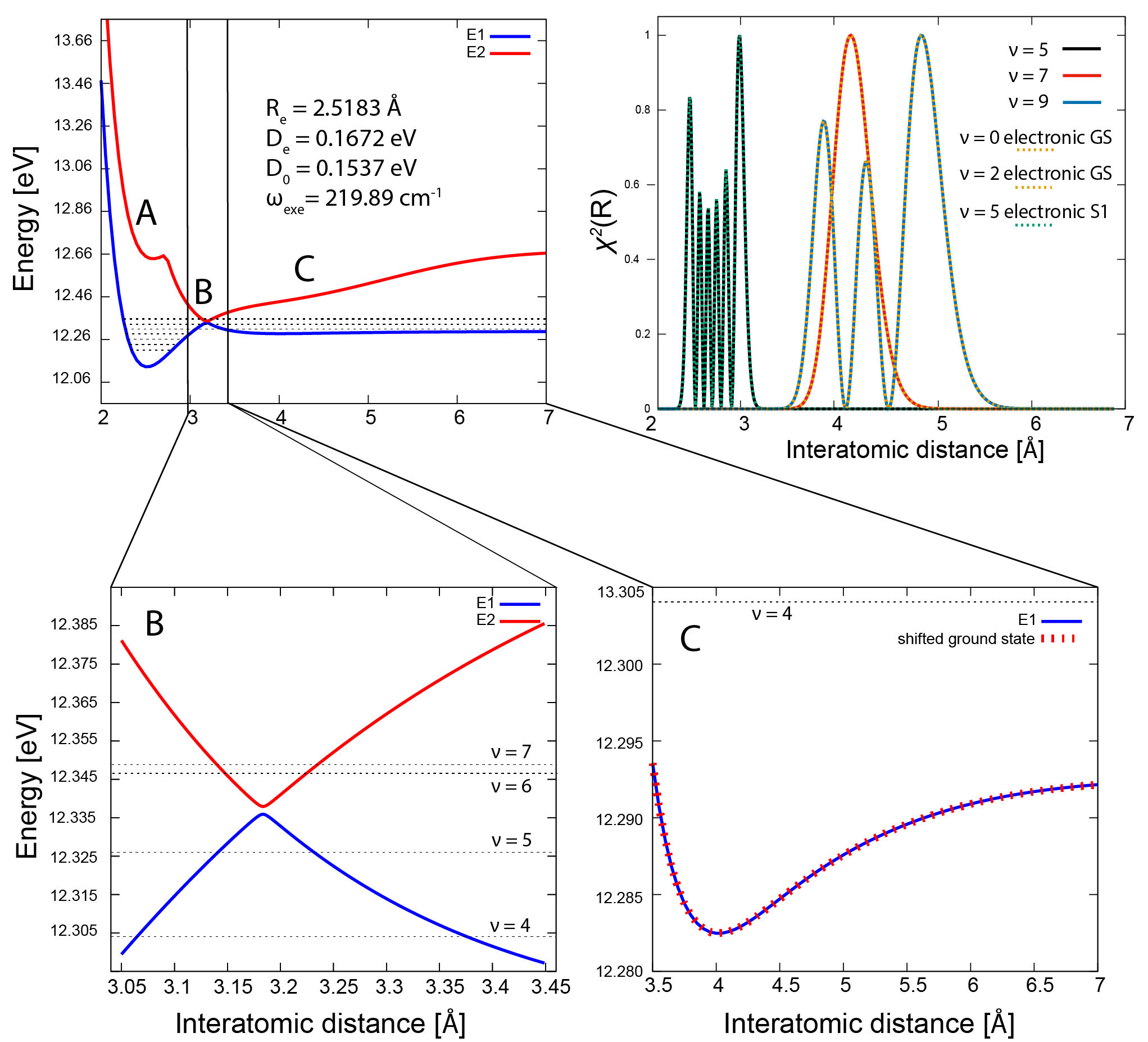}
    \caption{Potential energy surface (PES) for the first two excited states, E1 and E2, of the argon dimer using C-QED-CCSD-1-1 for a collective light-matter coupling strength $\lambda_0\sqrt{N}=$ \qty{0.04}{\atomicunit} (and $\lambda_0=$ \qty{0.0001}{\atomicunit}).
    The PES is partitioned into three regions defined by the avoided crossing distance $R_A$: region A for $R<R_A$, region B for $R\sim R_A$, and region C for $R>R_A$.
    Region B and C are highlighted, and the vibrational levels $\nu$ of E1 are indicated by horizontal dotted lines.
    Beyond the avoided crossing, the PES resembles the $\text{Ar}_2$ ground state PES, while in region A, it is the same as the electronic $\text{S}_1$ PES in \autoref{fig: Ar2 electronic}.
    We also report the square of the (unnormalized) vibrational wave functions $\chi^2(R)$ for selected vibrational levels.
    For the states, $\nu\leq 6$, the wave functions are similar to the $\text{S}_1$ vibrational levels. 
    Higher levels $\nu\ge 7$ resemble the ground state wave functions.
    Collective strong coupling thus induces an abrupt transition in the vibrational landscape of the first excited state.
    }
    \label{fig: Ar2 0.04 zoom}
\end{figure*}

We emphasize that the similarities between the C-QED-CC results in \autoref{fig: Ar2 excited states} and the electronic PESs of \autoref{fig: Ar2 electronic} do not imply that no modifications occur under collective strong coupling.
This is more clearly illustrated in \autoref{fig: Ar2 0.04 zoom}, where we increase the collective coupling strength to $\lambda_0\sqrt{N}=$ \qtylist{0.04}{\atomicunit} and highlight some features of the excited state PESs.
The ground state properties are fundamentally unchanged compared to the electronic simulation in \autoref{fig: Ar2 electronic}.
For the excited states, we partition the PESs in \autoref{fig: Ar2 0.04 zoom} into three regions.
Region A is restricted to the intermolecular distances below the avoided crossing $R<R_A$. 
Region B is defined by $R\sim R_A$.
Finally, region C is defined for intermolecular separations beyond the avoided crossing $R>R_A$.  
Even though the collective coupling is considerable, we see no changes in the harmonic properties of the first excited state E1 compared to the purely electronic case in \autoref{fig: Ar2 electronic}.
Therefore, the PES in region A is very similar to the electronic $\text{Ar}_2$ excited state PES.
In \autoref{fig: Ar2 0.04 zoom}, for region C, we plot the E1 PES and the ground state curve (shifted by the excitation energy of the lower polariton computed for an interatomic distance of $R=$ \qtylist{7.0}{\angstrom}).
The two curves are in almost perfect agreement, meaning that in region C, $\text{Ar}_2$ behaves as if it were in its ground state.
Region B can thus be interpreted as an avoided crossing between the electronic excited state of $\text{Ar}_2$ and its ground state shifted by the energy of the lower polariton. 
These states couple through the photonic degrees of freedom, and the resulting splitting depends on the microscopic light-matter coupling strength $\lambda_0$. 
In contrast, the position $R_A$ of the avoided crossing depends on the lower polariton energy and thus on the collective coupling $g$ (as well as on the details of the electronic ground and excited state PESs). \\
The PES E1 is then a combination of the $\text S_0$ and $\text S_1$ electronic surfaces, resulting in distinct vibrational properties compared to the non-polaritonic case.
In \autoref{fig: Ar2 0.04 zoom}, we plot the square of the (unnormalized) nuclear wave functions $\chi_\nu^2(R)$ of E1 for selected vibrational levels $\nu$. 
For comparison, we plot in dotted lines selected vibrational levels of $\text{S}_0$ and $\text{S}_1$ from \autoref{fig: Ar2 electronic}.
The first vibrational levels $\nu\leq 6$ of E1 are essentially identical to the $\text{S}_1$ vibrations.
In contrast, the states $\nu\geq 7$ resemble the ground state vibrational levels (though they also show very small oscillations in region A, see Sec. S3 of the Supporting Information). 
Therefore, collective strong coupling introduces an abrupt change in the vibrational landscape of the excited state, causing the higher vibrational states to behave like ground state vibrations.
The position $R_A$ and the energy $E_A$ of the avoided crossing are hence important since they determine the position of this sharp transition in the E1 vibrational ladder: the stronger the collective coupling, the shorter the distance $R_A$, and thus, the fewer levels $\nu$ behave like excimer vibrational states.
Therefore, once the coupling exceeds a critical value $g^*$, we can then expect a change in the vibrations involved in the Franck-Condon excitation process.
The system, initially confined in the $\text S_0$ vibrational levels, is then excited to ground state-like vibrational states. 
We therefore expect the excimer formation to be hindered since, while the $\text{S}_1$ PES and vibrational levels lead to the excimer, the $\text{Ar}$ atoms are less bound in the ground state.
Our results thus suggest that collective light-matter strong coupling inhibits the formation of excimers via a chemically different mechanism than for the single-molecule regime, for which we predict a weakening of the excimer bond due to a modified PES. 

\begin{figure}[!ht]
    \centering
    \includegraphics[width=.5\textwidth]{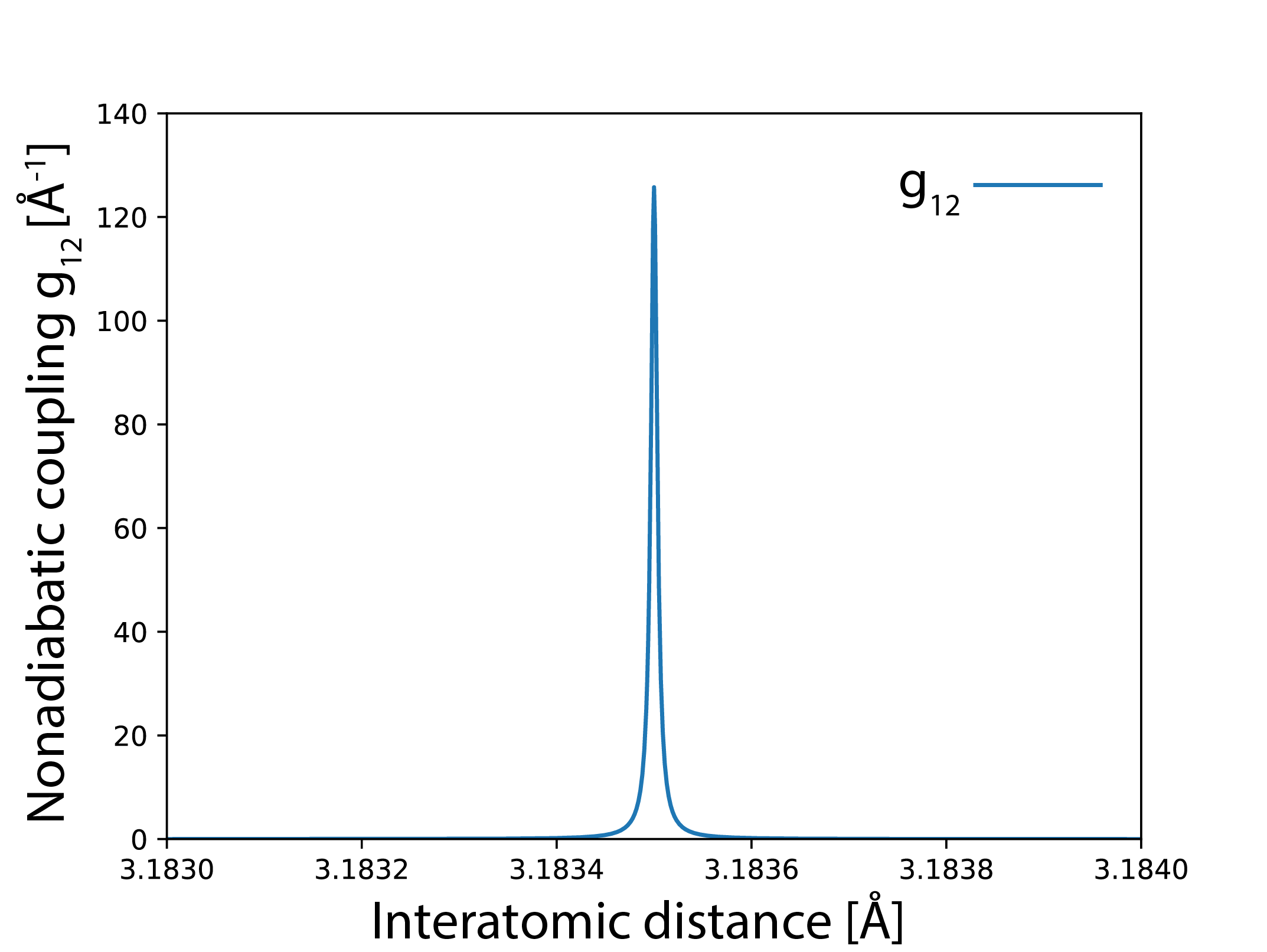}
    \caption{Nonadiabatic coupling $g_{12}$ computed using the Landau-Zener approximation for the argon dimer around the avoided crossing between E1 and E2 (see \autoref{fig: Ar2 0.04 zoom}).
    The nonadiabatic couplings are large but decay quickly far from the avoided crossing $R_A$.
    }
    \label{fig: nonadiab LANDAU}
\end{figure}

Since the excimer formation involves a motion from region C to region A, we now analyze the avoided crossing at $R_A$ in more detail.
This is relevant because several nuclear wave functions are non-negligible in region B and have energy comparable to or above the avoided crossing $E_A$.
In region B, there is a sharp change in the wave function character.
For $R>R_A$, the excitation E1 shows a delocalized (collective) polaritonic character, as indicated by the large $\gamma$ and $\theta$ amplitudes in the jacobian eigenvectors of \autoref{eq: jacobian}.
At the same time, E2 is fundamentally an excitation localized on the argon atoms, resembling the electronic excitation $\text{S}_1$.
For $R<R_A$, the characters of the wave functions are switched, as characteristic of avoided crossings. 
Since the energy splitting at $R_A$ is small ($\sim 10^{-3}$ \qty{}{\electronvolt}, determined by the microscopic light-matter coupling $\lambda_0$), we want to study the nonadiabatic couplings between E1 and E2.
As discussed above, region B can be described as an avoided crossing between $\text{S}_0$, shifted by the lower polariton energy, and $\text S_1$.
We now show that these states are strictly diabatic, and therefore, the Landau-Zener approximation is appropriate for computing nonadiabatic couplings.\cite{persico2018photochemistry}
Consider the uncoupled Hamiltonians
\begin{align}
    \tilde H_{e} & = \sum_i \frac{\bm{p}_i^2}{2}  +\sum_{i> j}\frac{1}{|\bm r_i - \bm r_j|} -\sum_{i,M}\frac{Z_M}{|\bm{r}_i-\bm{R}_M|} + \frac{1}{2} (\bm\lambda \cdot\bm d )^2 \label{eq: cbo ham}
\end{align}
\begin{align}
    \tilde H_{pol} & = \omega_o A^\dagger A + \omega b^\dagger b + g (b^\dagger + b) (A+A^\dagger)+ \frac{g^2}{\omega}(A+A^\dagger)^2 .\label{eq: hopfield ham}
\end{align}
The eigenfunctions of \autoref{eq: cbo ham}, where we include the dipole self-energy, can be obtained using coupled cluster cavity Born-Oppenheimer (CBO) method \cite{flick2017cavity, angelico2023coupled} implemented in the $e^{\mathcal{T}}$ program.\cite{folkestad2020t}
The Hamiltonian in \autoref{eq: hopfield ham}, which is independent of $R$, can be diagonalized by the Hopfield method \cite{hopfield1958theory}
\begin{equation}
    [H_{pol},\alpha_j^\dagger]=E_j \alpha_j^\dagger, \label{eq: hop method}
\end{equation}
where $\alpha_j^\dagger$ creates an excitation in the $j$-th (LP or UP) polariton mode of \autoref{eq: hopfield ham}.
Now consider the tensor product states
\begin{align}
    \ket{\tilde \eta_1} &= \ket{\widetilde {\text{S}_1}} \otimes \ket{0} \qquad
    \ket{\tilde {\eta}_2} = \ket{\widetilde {GS}} \otimes \alpha_{LP}^\dagger \ket{0},
\end{align}
where $\ket{\widetilde {\text{S}_1}}$ and $\ket{\widetilde {GS}}$ are the first excited state and the ground state of the Hamiltonian in \autoref{eq: cbo ham}, and $\ket{0}$ is the ground state (vacuum) the polariton modes in \autoref{eq: hop method}.
These states are strictly diabatic since they belong to different excitation spaces of the photon-dielectric Hamiltonian
\begin{equation}
    \braket{\tilde\eta_1|\frac{\partial}{\partial R}|\tilde\eta_2}=\braket{\tilde\eta_1|\frac{\partial^2}{\partial R^2}|\tilde\eta_2}=0,
\end{equation}
and are coupled by the bilinear and dipole-self energy terms in \autoref{eq: Ham final}
\begin{equation}
    V = \sqrt{\frac{\omega}{2}} (\bm{\lambda}\cdot \bm d ) (b+b^\dagger) + g \sqrt{\frac{2}{\omega}} (\bm{\lambda}\cdot \bm d ) (A+A^\dagger).
\end{equation}
We obtain the energy derivative $(\partial_R E_j)_{R_A}$ of $\ket{\tilde \eta_j}$ at the avoided crossing $R_A$ by finite difference of the CBO energies, and we compute the matrix element $V_{12}=\braket{\tilde\eta_1|V(R_A)|\tilde\eta_2}$ from the CBO transition dipole moments and the eigenfunctions of \autoref{eq: hopfield ham}.
Following the Landau-Zener procedure, the nonadiabatic coupling $g_{12}$ reads \cite{persico2018photochemistry}
\begin{equation}
    g_{12}(R-R_A) = \frac{-\beta/2}{1+\beta^2 (R-R_A)^2}, \label{eq: nonadiab coupling lorenz landau zener}
\end{equation}
where $\beta = \frac{(\partial_R E_1)_{R_A}-(\partial_R E_2)_{R_A}}{2\sqrt{V_{12}V_{21}}}\approx$ \qty{-250}{\per\angstrom} from our simulation of $\text{Ar}_2$.
The Lorentzian \autoref{eq: nonadiab coupling lorenz landau zener} is plotted in \autoref{fig: nonadiab LANDAU}.
The nonadiabatic coupling is non-negligible at the avoided crossing position $R_A$, though it decays quickly with the nuclear displacement $R-R_A$.
Already for a displacement of only \qty{0.0005}{\angstrom}, $g_{12}$ becomes virtually zero.
In addition, the rotational motion might also play a role in the vicinity of the avoided crossing.
Contrary to the standard electronic case, the optical device introduces an anisotropy in the system via its electromagnetic modes.\cite{castagnola2024polaritonic}
In our calculations, this is reflected in a dependence of the energies on the angle $\phi$ between the $\text{Ar}_2$ bond and the polarization $\boldsymbol{\epsilon}$.
Such dependence is small and can be neglected essentially everywhere except in region B.
In fact, the avoided crossing of \autoref{fig: Ar2 0.04 zoom} ($\phi=0$) becomes an intersection when $\text{Ar}_2$ is rotated to be perpendicular to $\boldsymbol{\epsilon}$ ($\phi=\pi/2$).
In Sec. S3 in the Supporting Information, we show using a configuration interaction singles (CIS) method that it is specifically a \textit{conical} intersection (the wave function changes sign traversing it).

A nuclear motion from region C to region A, corresponding to the excimer formation, thus involves a sharp change in the energy landscape, a sudden wave function localization, and an interplay of vibrational and rotational motions with nonnegligible nonadiabatic couplings around $R_A$.
We can thus expect a rich nuclear dynamics near the avoided crossing, and the details of the electronic PESs will presumably be relevant.

\section{Conclusions}\label{sec: conclusion}

In this paper, we develop and implement a quantum electrodynamics coupled cluster method for collective strong coupling based on the Hopfield description of polaritons.\cite{hopfield1958theory}
We discuss in detail the Hamiltonian and its physical interpretations as describing 1) an \textit{ab initio} molecule and $N$ harmonic toy-model molecules coupled to the photon field, 2) a molecule and an effective harmonic oscillator coupled to the photon field, and 3) a molecule and an infinite Drude dielectric coupled to the optical device.
We derive the Hartree-Fock and coupled cluster equations, which are implemented in $e^\mathcal{T}$,\cite{folkestad2020t} and simulate the properties of the argon excimer under light-matter strong coupling.

We find that QED-CCSD and our collective C-QED-CCSD-1-1 method provide different results.
For both models, the ground state properties are fundamentally unaltered by electronic strong coupling.
However, in the single-molecule regime (large $\lambda$), modeled using the QED-CC method, the hybridization of the $\text S_1$ state with the photon field leads to essential changes in the PESs.
This results in a weakening of the excimer bond, as evidenced by larger equilibrium distances and smaller harmonic frequencies.
On the other hand, under collective strong coupling, the predicted C-QED-CCSD-1-1 excimer PES can be described as merging the ground and excited state electronic PESs.
The excited state vibrational landscape is accordingly modified depending on the collective coupling strength.
The vibrational levels of the polaritonic excited state are separated into two groups: the lower-energy vibrations resembling the electronic excited state levels and the higher vibrations resembling the ground state vibrations.
We thus argue that once the collective coupling exceeds a critical value $g^*$, such that there is a change in the vibrational levels involved in the Frank-Condon excitation process, the excimer formation is effectively inhibited.
We also provide a detailed analysis of the avoided crossing region of the PES, where nonadiabatic couplings and wave function localization are relevant.
The presented discussion also shows some limitations of the single-molecule \textit{ab initio} QED methods, which could overestimate the effect of light-matter strong coupling if the coupling strength $\lambda$ is too large.
We suggest to have $\lambda\leq$ \qty{0.01}{\atomicunit} to avoid spurious effects.

Although we focused specifically on argon as a case study, $\text{Ar}_2$ provides a good and simple prototype for a broader discussion on excimer properties.
Excimers show PESs as those in \autoref{fig: Ar2 electronic}, and we expect the results discussed in the paper to have more general validity.
The described inhibition should then be a general trend, although the critical strength $g^*$ will change with the details of the system.
An atomic gas would be a good candidate for an experimental investigation since we avoid complications connected to inhomogeneous broadening, and the excimer has only one relevant nuclear coordinate (gas-phase platforms for molecular polaritons have recently been developed\cite{wright2023versatile, nelson2024more, chen2024exploring}).
Still, the high excitation energy is prohibitive for currently available experimental devices, and the argon transition is not very bright, which complicates achieving the strong coupling regime with large Rabi splittings.\cite{nelson2024more}
However, the system does not need to be in the gas phase.
For instance, nanofluidic cavities for electronic strong coupling have been developed,\cite{bahsoun2018electronic} and we believe they could be used to study excimers of organic molecules in solution.
The excitation energies and the oscillator strengths should then allow the strong and ultrastrong coupling regimes to be achieved more easily. 
Excimer-like systems can also be obtained in solid matrices.
An example is [2,2]-paracyclophane (PCP), which forms oriented crystals.\cite{haggag2024revisiting, diri2012electronic}
The two benzene rings of PCP get closer together upon excitation, showing a PES similar to \autoref{fig: Ar2 electronic}.
Experimentally, it could then be possible to, e.g., record a discontinuity in the emission properties of the system as a function of the collective coupling $g$, such that when $g>g^*$, the excimer emission should decrease. 
The vibrational structure of the absorption spectrum should also present a sharp change, though resolving the vibrational fine structure in a complicated electromagnetic environment such as a Fabry-Pérot cavity might be challenging.

We believe the developed method can provide important insight into polaritonic chemistry.
While we so far focused on properties under electronic strong coupling, future studies will be devoted to photochemistry and extending the presented formalism for vibrational strong coupling (VSC).

\section*{Supporting Information}
The supporting information includes further theoretical discussions and additional results.

\section*{Data avalability}

The $e^\mathcal{T}$ and VIBROT outputs are available in the following repository: \doi{10.5281/zenodo.13789200}.

\section*{Acknowledgements}
M.C. and H.K. acknowledge funding from the European Research Council (ERC) under the European Union’s Horizon 2020 Research and Innovation Programme (grant agreement No. 101020016).
M.C., M. L., and H. K. acknowledge funding from the Department of Chemistry at the Norwegian University of Science and Technology (NTNU).

\bibliography{main.bib}
\cleardoublepage

\begin{widetext}


\section*{Supplementary material}

# Supporting Information: Realistic ab initio predictions of excimer behavior under collective light-matter strong coupling

Matteo Castagnola, Marcus T. Lexander, and Henrik Koch*

*Department of Chemistry, Norwegian University of Science and Technology, 7491 Trondheim, Norway*

E-mail: henrik.koch@ntnu.no

# Contents

# S1 Theory

In this section, we provide further theoretical details on the methods used in the paper.

## S1.1 Hopfield theory of polaritons

In this section, we discuss the connection of our Hamiltonian with the theory of polaritons in linear dielectrics as presented by Hopfield[1].

The starting point is the Lagrangian for the electromagnetic field and a classical infinite dielectric

$$L = \int \mathrm{d}^2 \mathcal{L} \qquad (1)$$

with

$$\mathcal{L} = \frac{1}{8\pi}\left(\frac{1}{c}\frac{\partial \boldsymbol{A}}{\partial t} + \boldsymbol{\nabla}\varphi\right)^2 - \frac{1}{8\pi}(\boldsymbol{\nabla}\times\boldsymbol{A})^2 + \frac{1}{2\omega_0^2\beta}\left(\frac{\partial \boldsymbol{P}}{\partial t}\right)^2 - \frac{1}{2\beta}\boldsymbol{P}^2 + \varphi(\boldsymbol{\nabla}\cdot\boldsymbol{P}) + \frac{1}{c}\boldsymbol{A}\cdot\frac{\partial \boldsymbol{P}}{\partial t} \quad (2)$$

where $\boldsymbol{P}$ is a polarization density with a restoring force, and the electromagnetic field is described via the scalar $\varphi$ and vector potentials $\boldsymbol{A}$. The Euler-Lagrange equation of motion

$$\frac{\mathrm{d}}{\mathrm{d}t}\left(\frac{\partial \mathcal{L}}{\partial \dot{q}_i}\right) + \frac{\partial}{\partial_j}\frac{\partial \mathcal{L}}{\partial(\partial_j q_i)} = \frac{\partial \mathcal{L}}{\partial q_i} \quad (3)$$

from Equation 2 give

$$\frac{\partial^2 \boldsymbol{P}}{\partial t^2} + \omega_0^2 = \beta \boldsymbol{E} \quad (4)$$

where

$$\boldsymbol{E} = -\boldsymbol{\nabla}\varphi - \frac{1}{c}\frac{\partial \boldsymbol{A}}{\partial t}, \quad (5)$$

is the electric field. Classically, solving Equation 4 in the frequency domain results in a dielectric constant

$$\epsilon(\omega) = 1 + \frac{4\pi\beta\omega_0^2}{\omega_0^2 - \omega^2} \quad (6)$$

as in the Drude model[2]. The Lagrangian and Hamiltonian can then be expressed using the Fourier components of $\boldsymbol{A}$ and $\boldsymbol{P}$ and their conjugate momenta, imposing periodic boundary conditions

$$\boldsymbol{A} = \sum_{\boldsymbol{k},\lambda} c\sqrt{\frac{2\pi}{\omega_k V}}\boldsymbol{\epsilon}_\lambda(b_{\boldsymbol{k},\lambda}^\dagger e^{i\boldsymbol{k}\cdot\boldsymbol{r}} + b_{\boldsymbol{k},\lambda}e^{-i\boldsymbol{k}\cdot\boldsymbol{r}}) \quad (7)$$

$$\frac{\partial \mathcal{L}}{\partial \boldsymbol{A}} = \boldsymbol{\Pi} = i\sum_{\boldsymbol{k},\lambda}\frac{1}{c}\sqrt{\frac{\omega_k}{8\pi V}}\boldsymbol{\epsilon}_\lambda(b_{\boldsymbol{k},\lambda}^\dagger e^{i\boldsymbol{k}\cdot\boldsymbol{r}} - b_{\boldsymbol{k},\lambda}e^{-i\boldsymbol{k}\cdot\boldsymbol{r}}) \quad (8)$$

$$\boldsymbol{P} = \sum_{\boldsymbol{k},\lambda}\sqrt{\frac{\omega_0\beta}{2V}}\boldsymbol{\epsilon}_\lambda(A_{\boldsymbol{k},\lambda}^\dagger e^{i\boldsymbol{k}\cdot\boldsymbol{r}} + A_{\boldsymbol{k},\lambda}e^{-i\boldsymbol{k}\cdot\boldsymbol{r}}) \quad (9)$$

$$\frac{\partial \mathcal{L}}{\partial \boldsymbol{P}} = \boldsymbol{\Phi} = i\sum_{\boldsymbol{k},\lambda}\sqrt{\frac{1}{2\omega_0\beta V}}\boldsymbol{\epsilon}_\lambda(A_{\boldsymbol{k},\lambda}^\dagger e^{i\boldsymbol{k}\cdot\boldsymbol{r}} - A_{\boldsymbol{k},\lambda}e^{-i\boldsymbol{k}\cdot\boldsymbol{r}}). \quad (10)$$

where $a$ and $b$ are the annihilation operators for the polarization and the photon, respectively, and $V$ is the quantization volume. The Lagrangian in Equation 2 thus describes a set of oscillators coupled to the electromagnetic field, and the Hamiltonian reads

$$H' = \sum_{\boldsymbol{k},\lambda} \omega_k b^\dagger_{\boldsymbol{k},\lambda} b_{\boldsymbol{k},\lambda} + \omega_0 \sum_{\boldsymbol{k},\lambda} A^\dagger_{\boldsymbol{k},\lambda} A_{\boldsymbol{k},\lambda} + \sum_{\boldsymbol{k},\lambda} s_k (b^\dagger_{\boldsymbol{k},\lambda} + b_{-\boldsymbol{k},\lambda})(A^\dagger_{-\boldsymbol{k},\lambda} + A_{\boldsymbol{k},\lambda})$$
$$+ \sum_{\boldsymbol{k},\lambda} r_k (b^\dagger_{\boldsymbol{k},\lambda} + b_{-\boldsymbol{k},\lambda})(b^\dagger_{-\boldsymbol{k},\lambda} + b_{\boldsymbol{k},\lambda}), \tag{11}$$

with

$$s_k = \frac{\omega_0^2 \sqrt{\pi\beta}}{\sqrt{\omega_0 \omega_k}} \tag{12}$$

$$r_k = \frac{\pi\beta\omega_0^2}{\omega_k}. \tag{13}$$

The normal modes of the Hamiltonian, the polaritons, will be expressed in terms of new operators $\alpha_k$

$$\alpha_k = x a_k + y a_k^\dagger + z b_{-k} + w b^\dagger_{-k} \tag{14}$$

which can be found by solving the equation

$$[\alpha_k, H] = E_k \alpha_k \tag{15}$$

as described in Ref.[1].

Including also an external set of particles, the nonrelativistic lagrangian reads

$$L = \frac{1}{2} \sum_\alpha m_\alpha \dot{\boldsymbol{r}}_\alpha^2 + \int \mathrm{d}^3 r \, \mathcal{L} \tag{16}$$

where

$$\mathcal{L} = \frac{1}{8\pi}\left(\frac{1}{c}\frac{\partial \boldsymbol{A}}{\partial t} + \boldsymbol{\nabla}\varphi\right)^2 - \frac{1}{8\pi}(\boldsymbol{\nabla} \times \boldsymbol{A})^2 + \frac{1}{2\omega_0^2\beta}\left(\frac{\partial \boldsymbol{P}}{\partial t}\right)^2 - \frac{1}{2\beta}\boldsymbol{P}^2 + \varphi(\boldsymbol{\nabla} \cdot \boldsymbol{P}) + \frac{1}{c}\boldsymbol{A} \cdot \frac{\partial \boldsymbol{P}}{\partial t}$$
$$+ \boldsymbol{j} \cdot \boldsymbol{A}/c - \rho(\boldsymbol{r})\varphi(\boldsymbol{r}). \tag{17}$$

The equation of motion for the polarization density is unaltered, and in the Coulomb gauge $(\boldsymbol{\nabla} \cdot \boldsymbol{A}) = 0$, the equation for $\varphi$ is

$$\nabla^2 \varphi = -4\pi\rho(\boldsymbol{r}) + 4\pi(\boldsymbol{\nabla} \cdot \boldsymbol{P}), \tag{18}$$

which gives a Coulomb potential with external sources given by the external charges $\rho(\boldsymbol{r})$ and the medium polarization $\rho_{pol}(\boldsymbol{r}) \equiv -(\boldsymbol{\nabla} \cdot \boldsymbol{P})$, in agreement with the macroscopic Maxwell's equations in media. Since $\dot{\varphi}$ does not enter Equation 17, from Equation 18, we can eliminate $\varphi$ in favour of the other variables describing the systems, which is a necessary step for quantization. The equations for the particles and $\boldsymbol{A}$ give the Lorentz force and the Maxwell's equations for the vector potential, thus Equation 17 is a good lagrangian for our system. Using the Coulomb gauge and the definitions of the longitudinal and transverse electric fields, we can rewrite Equation 17 as

$$L = \frac{1}{2}\sum_\alpha m_\alpha \dot{\boldsymbol{r}}_\alpha^2 + \int \mathrm{d}^3 r\, \bar{\mathcal{L}} \tag{19}$$

where

$$\bar{\mathcal{L}} = \frac{1}{8\pi}\left(\frac{1}{c}\frac{\partial \boldsymbol{A}}{\partial t}\right)^2 - \frac{1}{8\pi}(\boldsymbol{\nabla} \times \boldsymbol{A})^2 + \frac{1}{2\omega_0^2\beta}\left(\frac{\partial \boldsymbol{P}}{\partial t}\right)^2 - \frac{1}{2\beta}\boldsymbol{P}^2 + \frac{1}{c}\boldsymbol{A} \cdot \frac{\partial \boldsymbol{P}}{\partial t} \tag{20}$$
$$+ \boldsymbol{j} \cdot \boldsymbol{A}/c - \rho(\boldsymbol{r})\varphi(\boldsymbol{r}) - \frac{1}{2}\varphi\rho + \frac{1}{2}\varphi(\boldsymbol{\nabla} \cdot \boldsymbol{P}). \tag{21}$$

The last two terms of Equation 21 are then the Coulomb interactions of the system (including the infinite self-interactions, which will be disregarded in there following).

For the Hamiltonian formalism, we need the conjugate momenta

$$p_i^\alpha = \frac{\partial L}{\partial \dot{r}_i^\alpha} = m_\alpha \dot{r}_i^\alpha + \frac{q_\alpha}{c} A_i(\boldsymbol{r}_\alpha) \tag{22}$$

$$\Pi_i = \frac{\partial \bar{\mathcal{L}}}{\partial \dot{A}_i} = \frac{1}{4\pi c^2} \dot{A}_i \tag{23}$$

$$\Phi_i = \frac{\partial \bar{\mathcal{L}}}{\partial \dot{P}_i} = \frac{1}{\omega_0^2 \beta} \dot{P}_i + \frac{A_i}{c}. \tag{24}$$

There is a clear analogy between the particles and the dielectric. The Hamiltonian then reads

$$H = \sum_\alpha \boldsymbol{p}_\alpha \cdot \dot{\boldsymbol{r}}_\alpha + \int d^3 r\, \boldsymbol{\Pi} \cdot \dot{\boldsymbol{A}} + \int d^3 r\, \boldsymbol{\Phi} \cdot \dot{\boldsymbol{P}} - L \tag{25}$$

$$= \sum_\alpha \frac{\left(\boldsymbol{p}_\alpha - \frac{q_\alpha}{c}\boldsymbol{A}(\boldsymbol{r}_\alpha)\right)^2}{2m_\alpha} + \frac{1}{2} \int d^3 r\, \varphi \left(\rho - (\boldsymbol{\nabla} \cdot \boldsymbol{P})\right)$$

$$+ \int d^3 r \left(2\pi c^2 \boldsymbol{\Pi}^2 + \frac{1}{8\pi}(\boldsymbol{\nabla} \times \boldsymbol{A})^2\right) + \int d^3 r \left(\frac{\omega_0^2 \beta}{2}\left(\boldsymbol{\Phi} - \frac{\boldsymbol{A}}{c}\right)^2 + \frac{1}{2\beta} \boldsymbol{P}^2\right). \tag{26}$$

Since we are interested in radiation modes, the longitudinal components of the polarization $\boldsymbol{P}$ are not interesting, and we can assume $\boldsymbol{P}$ to be purely transverse. This simplifies the analysis since $\boldsymbol{\nabla} \cdot \boldsymbol{P} = 0$ leads to the usual coulomb potential given by the particles only. We then use the Fourier expansion in a cube of volume $V$

$$\boldsymbol{A} = \sum_{\boldsymbol{k},\lambda} c \sqrt{\frac{2\pi}{\omega_k V}} \boldsymbol{\epsilon}_\lambda (b_{\boldsymbol{k},\lambda}^\dagger e^{i\boldsymbol{k}\cdot\boldsymbol{r}} + b_{\boldsymbol{k},\lambda} e^{-i\boldsymbol{k}\cdot\boldsymbol{r}}) \tag{27}$$

$$\boldsymbol{\Pi} = i \sum_{\boldsymbol{k},\lambda} \frac{1}{c} \sqrt{\frac{\omega_k}{8\pi V}} \boldsymbol{\epsilon}_\lambda (b_{\boldsymbol{k},\lambda}^\dagger e^{i\boldsymbol{k}\cdot\boldsymbol{r}} - b_{\boldsymbol{k},\lambda} e^{-i\boldsymbol{k}\cdot\boldsymbol{r}}) \tag{28}$$

$$\boldsymbol{P} = \sum_{\boldsymbol{k},\lambda} \sqrt{\frac{\omega_0 \beta}{2V}} \boldsymbol{\epsilon}_\lambda (A_{\boldsymbol{k},\lambda}^\dagger e^{i\boldsymbol{k}\cdot\boldsymbol{r}} + A_{\boldsymbol{k},\lambda} e^{-i\boldsymbol{k}\cdot\boldsymbol{r}}) \tag{29}$$

$$\boldsymbol{\Phi} = i \sum_{\boldsymbol{k},\lambda} \sqrt{\frac{1}{2\omega_0 \beta V}} \boldsymbol{\epsilon}_\lambda (A_{\boldsymbol{k},\lambda}^\dagger e^{i\boldsymbol{k}\cdot\boldsymbol{r}} - A_{\boldsymbol{k},\lambda} e^{-i\boldsymbol{k}\cdot\boldsymbol{r}}) \tag{30}$$

where $b$ and $b^\dagger$ are the annihilation and creation operators for the photon field, $A$ and $A^\dagger$ annihilate and create excitations in the environment oscillators, and $\boldsymbol{\epsilon}_\lambda$ $\lambda = 1, 2$ are the transverse polarizations of the field. We thus get

$$H = \sum_\alpha \frac{\left(\boldsymbol{p}_\alpha - \frac{q_\alpha}{c}\boldsymbol{A}(\boldsymbol{r}_\alpha)\right)^2}{2m_\alpha} + \frac{1}{2}\sum_{\alpha\neq\beta}\frac{q_\alpha q_\beta}{|\boldsymbol{r}_\alpha - \boldsymbol{r}_\beta|} + \sum_\alpha \varepsilon_\alpha$$
$$+ \sum_{\boldsymbol{k},\lambda} \omega_k b^\dagger_{\boldsymbol{k},\lambda} b_{\boldsymbol{k},\lambda} + \omega_0 \sum_{\boldsymbol{k},\lambda} A^\dagger_{\boldsymbol{k},\lambda} A_{\boldsymbol{k},\lambda} - i\sum_{\boldsymbol{k},\lambda} \frac{\omega_0^2 \sqrt{4\pi\beta}}{2\sqrt{\omega_0 \omega_k}}[A^\dagger_{\boldsymbol{k},\lambda} b^\dagger_{-\boldsymbol{k},\lambda} - A_{\boldsymbol{k},\lambda} b_{-\boldsymbol{k},\lambda} - A_{\boldsymbol{k},\lambda} b^\dagger_{\boldsymbol{k},\lambda} + A^\dagger_{\boldsymbol{k},\lambda} b_{\boldsymbol{k},\lambda}]$$
$$+ \sum_{\boldsymbol{k},\lambda} \frac{\pi\beta\omega_0^2}{\omega_k}[b^\dagger_{\boldsymbol{k},\lambda} b^\dagger_{-\boldsymbol{k},\lambda} + b_{\boldsymbol{k},\lambda} b_{-\boldsymbol{k},\lambda} + b^\dagger_{\boldsymbol{k},\lambda} b_{\boldsymbol{k},\lambda} + b_{\boldsymbol{k},\lambda} b^\dagger_{\boldsymbol{k},\lambda}]. \tag{31}$$

We can then rewrite the photon-dielectric and diamagnetic terms and use the rotation $A \to iA$ to get

$$H' = \sum_\alpha \frac{\left(\boldsymbol{p}_\alpha - \frac{q_\alpha}{c}\boldsymbol{A}(\boldsymbol{r}_\alpha)\right)^2}{2m_\alpha} + \frac{1}{2}\sum_{\alpha\neq\beta}\frac{q_\alpha q_\beta}{|\boldsymbol{r}_\alpha - \boldsymbol{r}_\beta|}$$
$$+ \sum_{\boldsymbol{k},\lambda} \omega_k b^\dagger_{\boldsymbol{k},\lambda} b_{\boldsymbol{k},\lambda} + \omega_0 \sum_{\boldsymbol{k},\lambda} A^\dagger_{\boldsymbol{k},\lambda} A_{\boldsymbol{k},\lambda} + \sum_{\boldsymbol{k},\lambda} \frac{\omega_0^2\sqrt{4\pi\beta}}{2\sqrt{\omega_0\omega_k}}(b^\dagger_{\boldsymbol{k},\lambda} + b_{-\boldsymbol{k},\lambda})(A^\dagger_{-\boldsymbol{k},\lambda} + A_{\boldsymbol{k},\lambda})$$
$$+ \sum_{\boldsymbol{k},\lambda} \frac{\pi\beta\omega_0^2}{\omega_k}(b^\dagger_{\boldsymbol{k},\lambda} + b_{-\boldsymbol{k},\lambda})(b^\dagger_{-\boldsymbol{k},\lambda} + b_{\boldsymbol{k},\lambda}) \tag{32}$$
$$\equiv \sum_\alpha \frac{\left(\boldsymbol{p}_\alpha - \frac{q_\alpha}{c}\boldsymbol{A}(\boldsymbol{r}_\alpha)\right)^2}{2m_\alpha} + \frac{1}{2}\sum_{\alpha\neq\beta}\frac{q_\alpha q_\beta}{|\boldsymbol{r}_\alpha - \boldsymbol{r}_\beta|}$$
$$+ \sum_{\boldsymbol{k},\lambda} \omega_k b^\dagger_{\boldsymbol{k},\lambda} b_{\boldsymbol{k},\lambda} + \omega_0 \sum_{\boldsymbol{k},\lambda} A^\dagger_{\boldsymbol{k},\lambda} A_{\boldsymbol{k},\lambda} + \sum_{\boldsymbol{k},\lambda} s_k (b^\dagger_{\boldsymbol{k},\lambda} + b_{-\boldsymbol{k},\lambda})(A^\dagger_{-\boldsymbol{k},\lambda} + A_{\boldsymbol{k},\lambda})$$
$$+ \sum_{\boldsymbol{k},\lambda} r_k (b^\dagger_{\boldsymbol{k},\lambda} + b_{-\boldsymbol{k},\lambda})(b^\dagger_{-\boldsymbol{k},\lambda} + b_{\boldsymbol{k},\lambda}) \tag{33}$$

where again

$$s_k = \frac{\omega_0^2 \sqrt{\pi \beta}}{\sqrt{\omega_0 \omega_k}} \tag{34}$$

$$r_k = \frac{\pi \beta \omega_0^2}{\omega_k}. \tag{35}$$

We then apply a transformation on the Hamiltonian

$$H \to e^{-iS} H e^{iS} \tag{36}$$

where

$$S = \int \mathrm{d}^3 r \, \boldsymbol{P} \cdot \frac{\boldsymbol{A}}{c}. \tag{37}$$

Using the relations

$$[A_j(\boldsymbol{r}), A_j(\boldsymbol{r}')] = 0 \tag{38}$$

$$[A_j(\boldsymbol{r}), \Pi_j(\boldsymbol{r}')] = i\delta^T_{ij}(\boldsymbol{r} - \boldsymbol{r}') \tag{39}$$

$$[P_j(\boldsymbol{r}), \Phi_j(\boldsymbol{r}')] = i\delta^T_{ij}(\boldsymbol{r} - \boldsymbol{r}') \tag{40}$$

we get the new Hamiltonian ($\boldsymbol{\nabla} \cdot \boldsymbol{P} = 0$)

$$\begin{aligned}\bar{H} = \sum_\alpha \frac{\left(\boldsymbol{p}_\alpha - \frac{q_\alpha}{c}\boldsymbol{A}(\boldsymbol{r}_\alpha)\right)^2}{2m_\alpha} + \frac{1}{2}\int \mathrm{d}^3r \varphi \rho + \int \mathrm{d}^3 r \left(2\pi c^2 \boldsymbol{\Pi}^2 + \frac{1}{8\pi}(\boldsymbol{\nabla} \times \boldsymbol{A})^2\right) \\ + \int \mathrm{d}^3 r \left(\frac{\omega_0^2 \beta}{2}\boldsymbol{\Phi}^2 + \frac{1}{2\beta}\boldsymbol{P}^2\right) + \int \mathrm{d}^3 r \, 2\pi \boldsymbol{P}^2 - 4\pi c \int \mathrm{d}^3 r \, \boldsymbol{\Pi} \cdot \boldsymbol{P}\end{aligned} \tag{41}$$

where we have a dipole self-energy term

$$\int d^3r\, 2\pi \boldsymbol{P}^2 = \pi\omega_0\beta \sum_{\boldsymbol{k},\lambda}[A^\dagger_{\boldsymbol{k},\lambda}A^\dagger_{-\boldsymbol{k},\lambda} + A_{\boldsymbol{k},\lambda}A_{-\boldsymbol{k},\lambda} + A^\dagger_{\boldsymbol{k},\lambda}A_{\boldsymbol{k},\lambda} + A_{\boldsymbol{k},\lambda}A^\dagger_{\boldsymbol{k},\lambda}] \quad (42)$$

$$= \pi\omega_0\beta \sum_{\boldsymbol{k},\lambda}(A^\dagger_{\boldsymbol{k},\lambda} + A_{-\boldsymbol{k},\lambda})(A^\dagger_{-\boldsymbol{k},\lambda} + A_{\boldsymbol{k},\lambda}) \quad (43)$$

and an interaction term

$$-4\pi c \int d^3r\, \boldsymbol{\Pi}\cdot\boldsymbol{P} = -i\sum_{\boldsymbol{k},\lambda}\sqrt{\pi\omega_0\omega_k\beta}[A^\dagger_{\boldsymbol{k},\lambda}b^\dagger_{-\boldsymbol{k},\lambda} - A_{\boldsymbol{k},\lambda}b_{-\boldsymbol{k},\lambda} + A_{\boldsymbol{k},\lambda}b^\dagger_{\boldsymbol{k},\lambda} - A^\dagger_{\boldsymbol{k},\lambda}b_{\boldsymbol{k},\lambda}] \quad (44)$$

$$= -i\sum_{\boldsymbol{k},\lambda}\sqrt{\pi\omega_0\omega_k\beta}(A^\dagger_{\boldsymbol{k},\lambda} + A_{-\boldsymbol{k},\lambda})(b^\dagger_{-\boldsymbol{k},\lambda} - b_{\boldsymbol{k},\lambda}). \quad (45)$$

We then work on the particles' degrees of freedom, assuming the dipole approximation by setting $A(\boldsymbol{r}_\alpha) \to \boldsymbol{A}(\boldsymbol{0})$. Then, we use the transformation

$$U = \exp\left(-i\sum_\alpha q_\alpha \boldsymbol{r}_\alpha \cdot \frac{\boldsymbol{A}(\boldsymbol{0})}{c}\right) \qquad H \to UHU^\dagger \quad (46)$$

to get

$$\sum_\alpha \frac{\left(\boldsymbol{p}_\alpha - \frac{q_\alpha}{c}\boldsymbol{A}(\boldsymbol{r}_\alpha)\right)^2}{2m_\alpha} \to \sum_\alpha \frac{\boldsymbol{p}_\alpha^2}{2m_\alpha} \quad (47)$$

$$\omega_k b^\dagger_{\boldsymbol{k},\lambda}b_{\boldsymbol{k},\lambda} \to \omega_k b^\dagger_{\boldsymbol{k},\lambda}b_{\boldsymbol{k},\lambda} + \frac{2\pi}{V}(\boldsymbol{\varepsilon}_{\boldsymbol{k},\lambda}\cdot\boldsymbol{d})^2 - i\sqrt{\frac{2\pi\omega_k}{V}}(\boldsymbol{\varepsilon}_{\boldsymbol{k},\lambda}\cdot\boldsymbol{d})(b_{\boldsymbol{k},\lambda} - b^\dagger_{\boldsymbol{k},\lambda}) \quad (48)$$

$$(b^\dagger_{-\boldsymbol{k},\lambda} - b_{\boldsymbol{k},\lambda}) \to (b^\dagger_{-\boldsymbol{k},\lambda} - b_{\boldsymbol{k},\lambda}) - i\sqrt{\frac{8\pi}{\omega_k V}}(\boldsymbol{\varepsilon}_{\boldsymbol{k},\lambda}\cdot\boldsymbol{d}), \quad (49)$$

where $\boldsymbol{d}$ is the dipole operator, so the new Hamiltonian reads

$$H = \sum_\alpha \frac{\boldsymbol{p}_\alpha^2}{2m_\alpha} + \frac{1}{2}\sum_{\alpha\neq\beta}\frac{q_\alpha q_\beta}{|\boldsymbol{r}_\alpha - \boldsymbol{r}_\beta|} + \sum_{\boldsymbol{k},\lambda}\omega_k b^\dagger_{\boldsymbol{k},\lambda}b_{\boldsymbol{k},\lambda} + \omega_0 \sum_{\boldsymbol{k},\lambda} A^\dagger_{\boldsymbol{k},\lambda}A_{\boldsymbol{k},\lambda} + \sum_{\boldsymbol{k},\lambda}\frac{2\pi}{V}(\boldsymbol{\varepsilon}_{\boldsymbol{k},\lambda}\cdot\boldsymbol{d})^2$$
$$+ \pi\omega_0\beta\sum_{\boldsymbol{k},\lambda}(A^\dagger_{\boldsymbol{k},\lambda}+A_{-\boldsymbol{k},\lambda})(A^\dagger_{-\boldsymbol{k},\lambda}+A_{\boldsymbol{k},\lambda}) - i\sum_{\boldsymbol{k},\lambda}\sqrt{\pi\omega_0\omega_k\beta}(A^\dagger_{\boldsymbol{k},\lambda}+A_{-\boldsymbol{k},\lambda})(b^\dagger_{-\boldsymbol{k},\lambda}-b_{\boldsymbol{k},\lambda})$$
$$- i\sqrt{\frac{2\pi\omega_k}{V}}(\boldsymbol{\varepsilon}_{\boldsymbol{k},\lambda}\cdot\boldsymbol{d})(b_{\boldsymbol{k},\lambda}-b^\dagger_{\boldsymbol{k},\lambda}) - \sum_{\boldsymbol{k},\lambda}\pi\sqrt{\frac{8\omega_0\beta}{V}}(A^\dagger_{\boldsymbol{k},\lambda}+A_{-\boldsymbol{k},\lambda})(\boldsymbol{\varepsilon}_{\boldsymbol{k},\lambda}\cdot\boldsymbol{d}). \quad (50)$$

Now we make use of the following rotations

$$b \to ib \qquad A \to -A \quad (51)$$

and

$$\sum_{\boldsymbol{k}} b_{-\boldsymbol{k}} = \sum_{\boldsymbol{k}} b_{\boldsymbol{k}}; \qquad \omega_k = \omega_{-k} \quad (52)$$

to write

$$H = \sum_\alpha \frac{\boldsymbol{p}_\alpha^2}{2m_\alpha} + \frac{1}{2}\sum_{\alpha\neq\beta}\frac{q_\alpha q_\beta}{|\boldsymbol{r}_\alpha - \boldsymbol{r}_\beta|} + \sum_{\boldsymbol{k},\lambda}\omega_k b^\dagger_{\boldsymbol{k},\lambda}b_{\boldsymbol{k},\lambda} + \omega_0 \sum_{\boldsymbol{k},\lambda} A^\dagger_{\boldsymbol{k},\lambda}A_{\boldsymbol{k},\lambda} + \sum_{\boldsymbol{k},\lambda}\frac{2\pi}{V}(\boldsymbol{\varepsilon}_{\boldsymbol{k},\lambda}\cdot\boldsymbol{d})^2$$
$$+ \pi\omega_0\beta\sum_{\boldsymbol{k},\lambda}(A^\dagger_{\boldsymbol{k},\lambda}+A_{-\boldsymbol{k},\lambda})(A^\dagger_{-\boldsymbol{k},\lambda}+A_{\boldsymbol{k},\lambda}) + \sum_{\boldsymbol{k},\lambda}\sqrt{\pi\omega_0\omega_k\beta}(A^\dagger_{\boldsymbol{k},\lambda}+A_{-\boldsymbol{k},\lambda})(b^\dagger_{-\boldsymbol{k},\lambda}+b_{\boldsymbol{k},\lambda})$$
$$\sqrt{\frac{2\pi\omega_k}{V}}(\boldsymbol{\varepsilon}_{\boldsymbol{k},\lambda}\cdot\boldsymbol{d})(b_{\boldsymbol{k},\lambda}+b^\dagger_{\boldsymbol{k},\lambda}) + \sum_{\boldsymbol{k},\lambda}\pi\sqrt{\frac{8\omega_0\beta}{V}}(A^\dagger_{\boldsymbol{k},\lambda}+A_{\boldsymbol{k},\lambda})(\boldsymbol{\varepsilon}_{\boldsymbol{k},\lambda}\cdot\boldsymbol{d}). \quad (53)$$

If we set

$$\pi\omega_0\beta \equiv \frac{g_{\boldsymbol{k}}^2}{\omega_k} \quad (54)$$

$$\frac{4\pi}{V} \equiv \lambda^2 \quad (55)$$

we have

$$\begin{aligned}H = &\sum_\alpha \frac{\boldsymbol{p}_\alpha^2}{2m_\alpha} + \frac{1}{2}\sum_{\alpha\neq\beta}\frac{q_\alpha q_\beta}{|\boldsymbol{r}_\alpha - \boldsymbol{r}_\beta|} + \sum_{\boldsymbol{k},\lambda}\omega_k b^\dagger_{\boldsymbol{k},\lambda}b_{\boldsymbol{k},\lambda} + \omega_0\sum_{\boldsymbol{k},\lambda}A^\dagger_{\boldsymbol{k},\lambda}A_{\boldsymbol{k},\lambda} + \sum_{\boldsymbol{k},\lambda}\frac{1}{2}(\boldsymbol{\lambda}_{\boldsymbol{k},\lambda}\cdot\boldsymbol{d})^2 \\ &+ \sum_{\boldsymbol{k},\lambda}\frac{g_{\boldsymbol{k}}^2}{\omega_k}(A^\dagger_{\boldsymbol{k},\lambda}+A_{-\boldsymbol{k},\lambda})(A^\dagger_{-\boldsymbol{k},\lambda}+A_{\boldsymbol{k},\lambda}) + \sum_{\boldsymbol{k},\lambda}g_{\boldsymbol{k}}(A^\dagger_{\boldsymbol{k},\lambda}+A_{-\boldsymbol{k},\lambda})(b^\dagger_{-\boldsymbol{k},\lambda}+b_{\boldsymbol{k},\lambda}) \\ &\sqrt{\frac{\omega_k}{2}}(\boldsymbol{\lambda}_{\boldsymbol{k},\lambda}\cdot\boldsymbol{d})(b_{\boldsymbol{k},\lambda}+b^\dagger_{\boldsymbol{k},\lambda}) + \sum_{\boldsymbol{k},\lambda}\sqrt{\frac{2}{\omega_k}}g_{\boldsymbol{k}}(A^\dagger_{\boldsymbol{k},\lambda}+A_{\boldsymbol{k},\lambda})(\boldsymbol{\lambda}_{\boldsymbol{k},\lambda}\cdot\boldsymbol{d}),\end{aligned} \quad (56)$$

which is analogous to a multimode description of the Hamiltonian in the main paper. We have then established a connection between the Hamiltonian in the main text and the Hamiltonian for an infinite (Drude) dielectric coupled to the electromagnetic field.

## S1.2 Collective QED Hartree-Fock

This section expands on the collective QED-HF equations presented in the main paper. The Hamiltonian is

$$\begin{aligned}H = &\sum_i \frac{\boldsymbol{p}_i^2}{2} + \frac{1}{2}\sum_{i\neq j}\frac{1}{|\boldsymbol{r}_i - \boldsymbol{r}_j|} - \sum_{i,M}\frac{Z_M}{|\boldsymbol{r}_i - \boldsymbol{R}_M|} + h_{nuc} \\ &+ \omega_0 A^\dagger A + \omega b^\dagger b \\ &+ g(b^\dagger + b)(A + A^\dagger) + \sqrt{\frac{\omega}{2}}(\boldsymbol{\lambda}\cdot\boldsymbol{d})(b + b^\dagger) \\ &+ \frac{1}{2}(\boldsymbol{\lambda}\cdot\boldsymbol{d})^2 + g\sqrt{\frac{2}{\omega}}(\boldsymbol{\lambda}\cdot\boldsymbol{d})(A + A^\dagger)\end{aligned} \quad (57)$$

where $\omega_0$ and $g$ can include the effect of the environment dipole self-energy as explained in the main text. The collective QED-HF model is then based on an uncorrelated ansatz wave function

$$|\text{R}\rangle = |\text{S}\rangle \otimes |P\rangle \otimes |E\rangle \quad (58)$$

where $|\text{S}\rangle$ is a Slater determinant of the molecule, $|P\rangle$ is a photon state, and $|E\rangle$ is an oscillator state. In its most general form, we write the photon and environment states in a

configuration interaction fashion

$$|P\rangle = \sum_n c_n \left(b^\dagger\right)^n |0\rangle \tag{59}$$

$$|E\rangle = \sum_n d_n \left(A^\dagger\right)^n |0\rangle. \tag{60}$$

In the first equation, $|0\rangle$ is the electromagnetic vacuum; in the second equation, $|0\rangle$ is the collective oscillator ground state; finally, $c_n$ and $d_n$ are general expansion coefficients. For any fixed Slater determinant and environment state, the photon state is obtained by minimization of the energy with respect to the averaged Hamiltonian

$$\langle S, E|H|S, E\rangle = \mathcal{K} + \omega b^\dagger b + g(b^\dagger + b)\langle E|A + A^\dagger|E\rangle + \sqrt{\frac{\omega}{2}}(\boldsymbol{\lambda} \cdot \langle S|\boldsymbol{d}|S\rangle)(b + b^\dagger) \tag{61}$$

where $\mathcal{K}$ is a constant that does not depend on the photon parameters. This Hamiltonian can diagonalized exactly with a coherent state transformation[3,4], and the minimum energy state is

$$|P\rangle = \exp\left(-\left[\frac{\boldsymbol{\lambda} \cdot \langle S|\boldsymbol{d}|S\rangle}{\sqrt{2\omega}} + \frac{g}{\omega}\langle E|A + A^\dagger|E\rangle\right](b^\dagger - b)\right)|0\rangle. \tag{62}$$

Using this state and a general Slater determinant, the minimum energy environment state can be obtained from the Hamiltonian

$$\langle S, P|H|S, P\rangle = \mathcal{K}' + \omega_0 A^\dagger A - 2\frac{g^2}{\omega}(A + A^\dagger)\langle E|A^\dagger + A|E\rangle \tag{63}$$

where $\mathcal{K}'$ does not include any oscillator operators $A$ or $A^\dagger$. This Hamiltonian is again diagonalized by a coherent state transformation, and we choose the environment state as

$$|E\rangle = \exp\left(\frac{2g^2}{\omega\omega_0}\langle E|A + A^\dagger|E\rangle(A^\dagger - A)\right)|0\rangle. \tag{64}$$

From this expression, we find

$$\langle E|A + A^\dagger|E\rangle = \frac{4g^2}{\omega\omega_0}\langle E|A + A^\dagger|E\rangle \implies \langle E|A + A^\dagger|E\rangle = 0. \tag{65}$$

The collective QED-HF state is thus

$$|\mathrm{R}\rangle = |\mathrm{HF}\rangle \otimes \exp\left(-\frac{\boldsymbol{\lambda}\cdot\langle\boldsymbol{d}\rangle_{\mathrm{HF}}}{\sqrt{2\omega}}(b^\dagger - b)\right)|0\rangle \otimes |0\rangle \tag{66}$$

$$\equiv U_{\mathrm{QED\text{-}HF}}|\mathrm{HF}, 0, 0\rangle. \tag{67}$$

Transforming the Hamiltonian using $U_{\mathrm{QED\text{-}HF}}$ we obtain

$$\begin{aligned}U_{\mathrm{QED\text{-}HF}}^\dagger H\, U_{\mathrm{QED\text{-}HF}} = \\ \sum_i \frac{\boldsymbol{p}_i^2}{2} + \frac{1}{2}\sum_{i\neq j}\frac{1}{|\boldsymbol{r}_i - \boldsymbol{r}_j|} - \sum_{i,M}\frac{Z_M}{|\boldsymbol{r}_i - \boldsymbol{R}_M|} + h_{nuc} \\ + \frac{1}{2}(\boldsymbol{\lambda}\cdot(\boldsymbol{d} - \langle\boldsymbol{d}\rangle_{\mathrm{QED\text{-}HF}}))^2 + \tilde{\omega}_0 A^\dagger A + \omega b^\dagger b \\ + \tilde{g}(b^\dagger + b)(A + A^\dagger) + \sqrt{\frac{\omega}{2}}(\boldsymbol{\lambda}\cdot(\boldsymbol{d} - \langle\boldsymbol{d}\rangle_{\mathrm{QED\text{-}HF}}))(b + b^\dagger) \\ + \tilde{g}\sqrt{\frac{2}{\omega}}(\boldsymbol{\lambda}\cdot(\boldsymbol{d} - \langle\boldsymbol{d}\rangle_{\mathrm{QED\text{-}HF}}))(A + A^\dagger)\end{aligned} \tag{68}$$

which is manifestly origin invariant. The Slater determinant is then obtained from the effective Hamiltonian

$$\langle 0, 0|U_{\mathrm{QED\text{-}HF}}^\dagger H\, U_{\mathrm{QED\text{-}HF}}|0, 0\rangle, \tag{69}$$

which gives the same Fock matrix as in QED-HF[4].

### S1.2.1 Response equations

The response equations for the collective QED-HF model are obtained from the time-dependent ansatz wave function

$$\exp\big(-i\Lambda(t)\big)|\text{R}\rangle = \exp\big(-i\Lambda(t)\big)\,|\text{HF}\rangle \otimes |0\rangle \otimes |0\rangle, \tag{70}$$

where $\Lambda(t)$ is the Hermitian operator

$$\begin{aligned}\Lambda(t) &= \frac{1}{\sqrt{2}}\sum_{ai}\big(\kappa_{ai}E_{ai} + \kappa_{ai}^{*}E_{ia}\big) + \sum\big(\gamma b^{\dagger} + \gamma^{*}b\big) + \big(\Theta A^{\dagger} + \Theta^{*}A\big) \\ &= \boldsymbol{\kappa}(t) + \sum\big(\gamma b^{\dagger} + \gamma^{*}b\big) + \big(\Theta A^{\dagger} + \Theta^{*}A\big)\end{aligned} \tag{71}$$

in the QED-HF transformed representation of the Hamiltonian of Equation 68 subject to an external perturbation $V^{t}$. The parameterization is the same as TD-QED-HF[5] but with the additional time-evolution coherent-state operator for the environment defined by the parameter $\Theta$. Following the same procedure as HF response theory, performing a Fourier transform and using Ehrenfest's theorem, we then get the response equation

$$\left[\begin{pmatrix}\mathbf{A} & \mathbf{B} \\ \mathbf{B}^{*} & \mathbf{A}^{*}\end{pmatrix} - \omega\begin{pmatrix}\mathbf{1} & 0 \\ 0 & -\mathbf{1}\end{pmatrix}\right]\begin{pmatrix}\boldsymbol{X} \\ \boldsymbol{Y}\end{pmatrix} = i\begin{pmatrix}\boldsymbol{g}_1 \\ \boldsymbol{g}_2\end{pmatrix} \equiv i\boldsymbol{g}, \tag{72}$$

where the vectors $\boldsymbol{X}$ and $\boldsymbol{Y}$ collect the Fourier transformed parameters

$$\boldsymbol{X} = \begin{pmatrix}\gamma^{\Omega} \\ \kappa_{ai}^{\Omega} \\ \Theta^{\Omega}\end{pmatrix} \quad \boldsymbol{Y} = \begin{pmatrix}[\gamma^{-\Omega}]^{*} \\ [\kappa_{ai}^{-\Omega}]^{*} \\ [\Theta^{-\Omega}]^{*}\end{pmatrix}. \tag{73}$$

corresponding to the (external) frequency $\Omega$. The right-hand side of Equation 70 is the generalized gradient

$$\boldsymbol{g}_1 = \begin{pmatrix} \langle [b, V^\Omega] \rangle_\text{R} \\ \frac{1}{\sqrt{2}} \langle [E_{ia}, V^\Omega] \rangle_\text{R} \\ \langle [A, V^\Omega] \rangle_\text{R} \end{pmatrix} \quad \boldsymbol{g}_2 = \begin{pmatrix} \langle [b^\dagger, V^\Omega] \rangle_\text{R} \\ \frac{1}{\sqrt{2}} \langle [E_{ai}, V^\Omega] \rangle_\text{R} \\ \langle [A^\dagger, V^\Omega] \rangle_\text{R} \end{pmatrix}. \tag{74}$$

The matrix

$$\boldsymbol{E}^{[2]} = \begin{pmatrix} \boldsymbol{A} & \boldsymbol{B} \\ \boldsymbol{B}^* & \boldsymbol{A}^* \end{pmatrix} \tag{75}$$

is the generalized Hessian matrix, while

$$\boldsymbol{S}^{[2]} = \begin{pmatrix} \boldsymbol{1} & 0 \\ 0 & -\boldsymbol{1} \end{pmatrix} \tag{76}$$

is the generalized metric matrix. The excitation energies and transition properties are obtained from the pseudo-eigenvalue equation

$$\left[\begin{pmatrix} \boldsymbol{A} & \boldsymbol{B} \\ \boldsymbol{B}^* & \boldsymbol{A}^* \end{pmatrix} - \omega \begin{pmatrix} \boldsymbol{1} & 0 \\ 0 & -\boldsymbol{1} \end{pmatrix}\right] \begin{pmatrix} \boldsymbol{X} \\ \boldsymbol{Y} \end{pmatrix} = i \begin{pmatrix} \boldsymbol{0} \\ \boldsymbol{0} \end{pmatrix} \tag{77}$$

where

$$\boldsymbol{A} = \begin{pmatrix} \langle [b, [H, b^\dagger]] \rangle_\text{R} & \frac{1}{\sqrt{2}} \langle [b, [H, E_{ai}]] \rangle_\text{R} & \langle [A, [H, b^\dagger]] \rangle_\text{R} \\ \frac{1}{\sqrt{2}} \langle [b^\dagger, [H, E_{ia}]] \rangle_\text{R} & \frac{1}{2} \langle [E_{jb}, [H, E_{ai}]] \rangle_\text{R} & \frac{1}{\sqrt{2}} \langle [A, [H, E_{ai}]] \rangle_\text{R} \\ \langle [A^\dagger, [H, b]] \rangle_\text{R} & \frac{1}{\sqrt{2}} \langle [A^\dagger, [H, E_{ia}]] \rangle_\text{R} & \langle [A, [H, A^\dagger]] \rangle_\text{R} \end{pmatrix} \tag{78}$$

$$\boldsymbol{B} = \begin{pmatrix} \langle [b, [H, b]] \rangle_\text{R} & \frac{1}{\sqrt{2}} \langle [b, [H, E_{ia}]] \rangle_\text{R} & \langle [A, [H, b]] \rangle_\text{R} \\ \frac{1}{\sqrt{2}} \langle [b^\dagger, [H, E_{ai}]] \rangle_\text{R} & \frac{1}{2} \langle [E_{bj}, [H, E_{ai}]] \rangle_\text{R} & \frac{1}{\sqrt{2}} \langle [A, [H, E_{ia}]] \rangle_\text{R} \\ \langle [A^\dagger, [H, b^\dagger]] \rangle_\text{R} & \frac{1}{\sqrt{2}} \langle [A^\dagger, [H, E_{ai}]] \rangle_\text{R} & \langle [A^\dagger, [H, A^\dagger]] \rangle_\text{R} \end{pmatrix}. \tag{79}$$

If we disregard the matrix **B**, we get the so-called Tamm-Dancoff approximation, equivalent to a CIS model.

## S1.3 Collective QED Coupled Cluster

This section expands on the collective QED coupled cluster equations presented in the main text.

The CC ansatz is defined in the coherent state-transformed representation of the Hamiltonian, here expressed in second quantization

$$
\begin{aligned}
H = &\sum_{pq} h_{pq} E_{pq} + \frac{1}{2}\sum_{pqrs} g_{pqrs} e_{pqrs} + \frac{1}{2}\sum_{pqrs}(\boldsymbol{\lambda}\cdot(\boldsymbol{d}-\langle\boldsymbol{d}\rangle_{\text{QED-HF}}))_{pq}(\boldsymbol{\lambda}\cdot(\boldsymbol{d}-\langle\boldsymbol{d}\rangle_{\text{QED-HF}}))_{rs} E_{pq} E_{rs} \\
&+ \omega b^\dagger b + \mathcal{E} A^\dagger \bar{A} + \sqrt{\frac{\omega}{2}}\sum_{pq}(\boldsymbol{\lambda}\cdot(\boldsymbol{d}-\langle\boldsymbol{d}\rangle_{\text{QED-HF}}))_{pq} E_{pq}(b^\dagger+b) + g(A+A^\dagger)(b+b^\dagger) \\
&+ \frac{\bar{g}^2}{\omega}(A+A^\dagger)^2 + \sqrt{\frac{2}{\omega}} g (A+A^\dagger)\sum_{pq}(\boldsymbol{\lambda}\cdot(\boldsymbol{d}-\langle\boldsymbol{d}\rangle_{\text{QED-HF}}))_{pq} E_{pq},
\end{aligned}
\tag{80}
$$

where $E_{pq}$ and $e_{pqrs}$ are the spin adapted singlet operators[6] in an orthonormal orbital basis indexed with $p, q, r$, and $s$. The wave function reads

$$|\text{CC}\rangle = e^T |\text{HF}, 0, 0\rangle, \tag{81}$$

where $T$ is the cluster operator

$$T = \sum_\mu t_\mu \tau_\mu \tag{82}$$

where $t_\mu$ are the cluster amplitudes associated with the excitation operator $\tau_\mu$, which creates excited states determinant as well as photon and environment excitations $\tau_\mu |HF, 0, 0\rangle = |\mu\rangle$. We implemented the QED-CCSD-1-1 method, in which $T$ has up to double excitation for

the electrons and single excitations for the photons and the environment

$$T = \sum_{ai} t_{ai} E_{ai} + \frac{1}{2} \sum_{aibj} t_{aibj} E_{ai} E_{bj}$$
$$+ \sum_{ai} s_{ai} E_{ai} b^\dagger + \frac{1}{2} \sum_{aibj} s_{aibj} E_{ai} E_{bj} b^\dagger$$
$$+ \gamma b^\dagger + \theta A^\dagger + \kappa A^\dagger b^\dagger$$
$$+ \sum_{ai} o_{ai} E_{ai} A^\dagger + \frac{1}{2} \sum_{aibj} o_{aibj} E_{ai} E_{bj} A^\dagger$$
$$+ \sum_{ai} \rho_{ai} E_{ai} A^\dagger b^\dagger + \frac{1}{2} \sum_{aibj} \rho_{aibj} E_{ai} E_{bj} A^\dagger b^\dagger \qquad (83)$$

where $i$ and $j$ ($a$ and $b$) label occupied (virtual) orbitals and $E_{ai}$ are the excitation singlet operators. The ground state amplitudes are obtained by projecting the Schrödinger equation

$$H \ket{CC} = E \ket{CC} \qquad (84)$$

onto the subspace $S$, including the HF reference and all the excitations in $T$

$$S = \{\ket{HF,0,0}, \ket{HF,1,0}, \ket{HF,0,1}, \ket{HF,1,1}, E_{ai}\ket{HF,0,0}, \ldots, E_{ai}E_{bj}\ket{HF,1,1}\}. \qquad (85)$$

The energy is obtained by projection onto the HF reference state

$$E_{CC} = \bra{HF,0,0} H \ket{CC} \qquad (86)$$

and the amplitudes are obtained by solving the equations

$$\Omega_\mu = \bra{HF,0,0} \tau_\mu^\dagger e^{-T} H e^T \ket{HF,0,0} = 0. \qquad (87)$$

Following the coupled cluster theory[6], we can define the CC dual state

$$\langle \Lambda | = \langle HF, 0, 0 | + \sum_\mu \bar{t}_\mu \langle \mu | \tag{88}$$

where $\bar{t}_\mu$ are the Lagrange multipliers obtained by solving the equation

$$\bar{\boldsymbol{t}}^T \boldsymbol{A} = -\boldsymbol{\eta}^T \tag{89}$$

where

$$\eta_\mu = \langle HF, 0, 0 | e^{-T} [H, \tau_\mu] e^T | HF, 0, 0 \rangle \tag{90}$$

and

$$A_{\mu,\nu} = \langle \mu | e^{-T} [H, \tau_\nu] e^T | R \rangle \tag{91}$$

is the CC Jacobian. The ground state properties (e.g. the dipole moment) are then obtained from the one-electron density matrix

$$D_{pq} = \langle \Lambda | E_{pq} | CC \rangle . \tag{92}$$

For the excited states, we implemented the equation of motion (EOM) CC equations[4,6]. The excited states are expressed in a configuration interaction fashion

$$|E\rangle = e^T \sum_\mu c_\mu |\mu\rangle \tag{93}$$

$$\langle E | = \sum_\mu \bar{c}_\mu \langle \mu | e^{-T} \tag{94}$$

and the CI vector $(c_\mu)$ is found from the left and right eigenvalue equations

$$\boldsymbol{A}\boldsymbol{c} = E\boldsymbol{c} \tag{95}$$

$$\bar{\boldsymbol{c}}^T \boldsymbol{A} = E\bar{\boldsymbol{c}}^T . \tag{96}$$

The eigenvalues of the Jacobian $\boldsymbol{A}$ then give the excitation energies of the system, and the left and right eigenvectors (which are different since CC is a non-hermitian theory) provide the states. The state and transition properties are then obtained from the one electron densities

$$D_{pq}^{12} = \langle 1|E_{pq}|2\rangle. \tag{97}$$

## S2 Computational details

The QED-CCSD-1-1 equations are implemented in a private branch of the $e^{\mathcal{T}}$ program, an open-source electronic structure program[7]. We also implemented linear response for the collective QED-HF model (collective TD-QED-HF and configuration interaction singles (CIS or Tamm-Dancoff)). The Cavity Born-Oppenheimer (CBO) and electronic calculations are computed using the development branch of $e^{\mathcal{T}}$, and the vibrational analysis of the potential energy surfaces is computed with the VIBROT module of the OpenMolcas program[8].

We reproduced the results in Ref.[3] to validate the proposed QED-CCSD-1-1 wave function. The system under study is a cluster of 5 hydrogen molecules $(H_2)_5$ with an intermolecular distance of $5\,\text{Å}$ in an H-aggregate configuration. The central molecule is labelled P, and it has a slightly longer bond length ($0.78\,\text{Å}$) than the other $H_2$ ($0.76\,\text{Å}$), and the first and second adjacent $H_2$ are labelled A1 and A2, respectively. In addition, $N$ hydrogen molecules ($N = 1 - 6$) are added at a distance of $20\,\text{Å}$. The intermolecular forces are relevant in the central $(H_2)_5$ cluster and directly couple the hydrogens. At the same time, the molecules at a larger distance are only indirectly coupled via the QED photon field. The photon field polarization is along the $H_2$ bond, with a light-matter coupling strength of $\lambda = 0.005$ a.u. and tuned to the first excitation of a single $H_2$ molecule computed at CCSD level of theory $\omega_p = 12.498\,\text{eV}$. The excitation energies and oscillator strengths are computed using QED-CCSD[4] using aug-cc-pVDZ as basis set. The same calculations are carried out for QED-CCSD-1-1, retaining only the central $(H_2)_5$ cluster, while the farther away molecules are approximated

by toy-model harmonic molecules as explained in the main text. The harmonic energy and oscillator strengths are obtained from the electronic CCSD simulation of a single $H_2$. The state $\rho_k$ and transition $\rho_{j0}$ identities are saved in cube files with a grid spacing of $0.1\,\text{Å}$. The lower polariton left and right transition densities are then integrated around the hydrogens in the central cluster and perpendicularly to the $H_2$ bond

$$\varrho_{j0}^I(x) = \int_{-\infty}^{+\infty} dz \int_{Y_I-2.0\,\text{Å}}^{Y_I+2.0\,\text{Å}} dy\, \rho_{j0}(x,y,z) \quad I = \text{P, A1, A2} \tag{98}$$

with a buffer of $2\,\text{Å}$ around each $H_2$.

The argon excimer $Ar_2$ energies and properties are computed using the aug-cc-pVDZ basis set. For the QED-CCSD results, we employed a light matter coupling strength of $\lambda = 0.01$ a.u. and $0.025$ a.u., while for the QED-CCSD-1-1 results, we set $\lambda_0 = 0.0001$ a.u. and $N$ such that $\lambda_0\sqrt{N} = 0.01$ a.u. and $0.025$ a.u.; that is, the collective coupling is the same as the single molecule coupling of the QED-CCSD simulations. The field polarization is along the $Ar_2$ bond, and the photon is tuned to the first bright excitation of a single argon atom $\omega_p = 12.67$ eV. The energy surfaces are calculated with a step of $0.005\,\text{Å}$ for the intermolecular separation of the argon atoms.

For the cavity Born-Oppenheimer simulations, we set $\lambda = 0.0001$ a.u. and field strength parameter $q = 0$ to only include the dipole self-energy in the Hamiltonian.

For the CIS calculations, we employ the aug-cc-pVDZ basis set and locate the conical intersection by setting the field polarization perpendicular to the argon bond. We then circulate the intersection starting with a slightly stretched bong length and $Ar_2$ with an angle $\phi = 60°$ with the field polarization. We first rotated the bond until $\phi = 120°$, compressed the bond until a length shorter than the intersection bong length, then rotated back the system until $\phi = 60°$, and stretched the bong length again to the initial value. The CIS coefficients are recorded for every step of the circulation.

The relevant molecular geometries are reported in section S4. All the relevant outputs

are available in the repository `DOI:10.5281/zenodo.13789200`.

# S3 Additional results

In this section, we provide additional results complementing the ones presented in the paper.

## S3.1 Additional results for the hydrogen cluster – wave function validation

In the paper, we showed that we could reproduce the results of Ref.[3] by substituting the molecules far away from the central $(H_2)_5$ cluster with harmonic oscillators. We, therefore, used the same parameters as Ref.[3], that is $\lambda = 0.005$ a.u. adding $N = 1-6$ molecules (QED-CCSD) or harmonic oscillators (QED-CCSD-1-1). In Figure S1, we show the lower polariton left integrated transition densities obtained by using $\lambda_0 = 0.00001$ a.u. and $\tilde{N}$ such that $\lambda_0\sqrt{\tilde{N}} = \lambda\sqrt{N}$. The figure shows that the qualitative sign change of the local properties

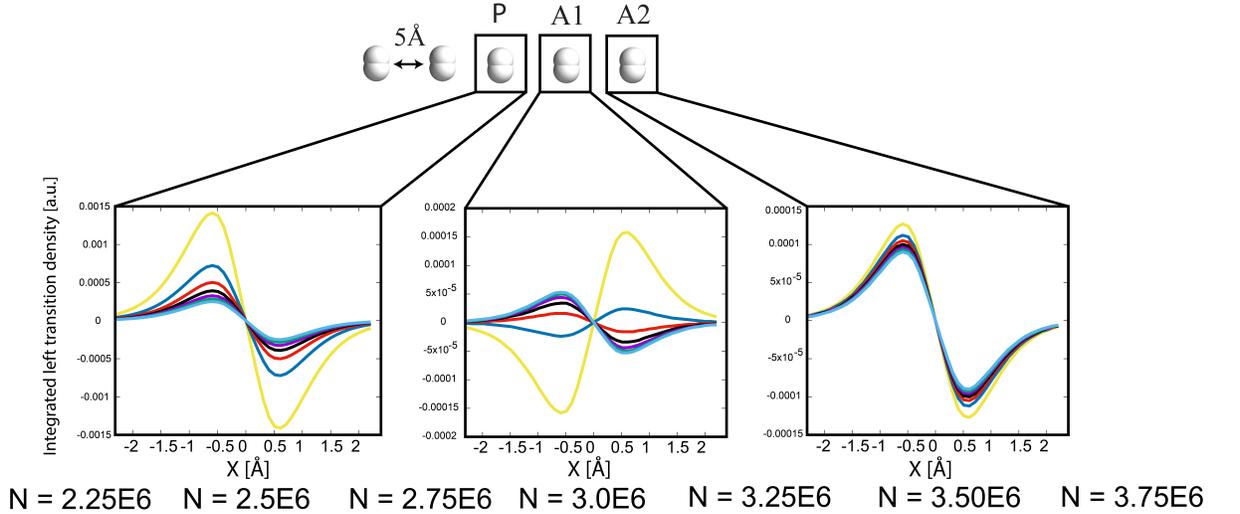

Figure S1: Integrated local left transition densities for the hydrogens P (perturbed) A1 and A2 (first and second closest to P) for a low coupling strength $\lambda_0 = 0.00001$ a.u. and large number of toy model harmonic molecules $N$ computed at QED-CCSD-1-1 level of theory.

of A1 is maintained even under collective strong coupling, for small light-matter coupling strengths and large number of molecules. At the same time, compared to the results in

Ref[3] and in the main paper, the scale of the P local transition density is much lower due to the larger delocalization of the wave function. Nevertheless, the contribution of P is still larger than the contribution of A1 and A2. In Ref.[3], care is taken in studying the so-called "thermodynamic limit". That is, since we make predictions based on relatively large coupling strengths and only a few molecules, we need to ensure that the results also make sense for small $\lambda$ and large $N$. The thermodynamic limit is defined by the equation

$$N \to \infty, \qquad \lambda\sqrt{N} \to \text{constant.} \tag{99}$$

In Figure S1 we see that this is the case by decreasing the coupling strength and increasing the number of environment $H_2$. Another way to achieve this limit is to increase both the number of environment oscillators and the number of clusters $(H_2)_5$ (or perturbed hydrogens P). This could be done, e.g., by adding more effective oscillators in the Hamiltonian to mimic the excitation energies of the cluster. However, we follow the same strategy of Ref.[3] and explicitly add more and more replicas $N_{rep}$ of the $(H_2)_5$ cluster. In Ref.[3], only two replicas of the system were considered because of the steep computational demands of the coupled cluster methods. Here, since we do not need to include explicitly the far-away hydrogens, replaced by effective oscillators, we can have more replicas $N_{rep}$. In Figure S2, we show the local transition density for A1 for $N_{rep} = 2 - 5$, rescaled by $\sqrt{N_{rep}}$. The light matter coupling strength is set to $\lambda/\sqrt{N_{rep}}$, $\lambda = 0.005$ a.u. as in the previous calculations. For each $N_{rep}$ simulation, we included via the QED-CCSD-1-1 method a number of fictitious harmonic toy molecules given by $N_{rep}N$, where $N = 3 - 6$. Figure S2 shows that the local contribution of each molecule maintained the same qualitative behaviour with the collective coupling strength as the number of replicas increases, and the contribution scales down with $\sqrt{N_{rep}}$. This is the same claim made in Ref.[3], which shows that the contribution scales down as the number of perturbed molecules, not the total number of molecules in the simulation.

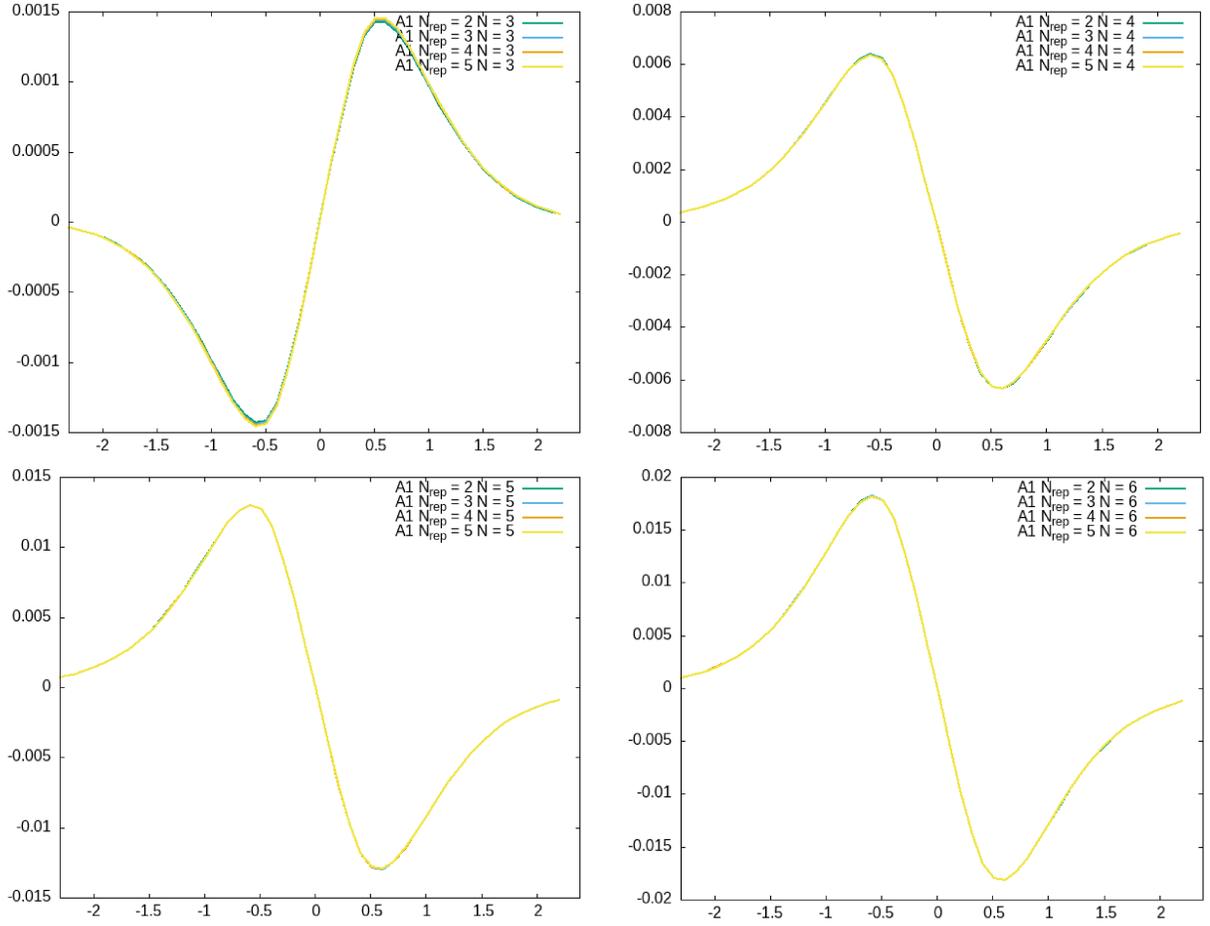

Figure S2: Local integrated transition density for A1 computed with $N_{rep}$ explicit replicas of the $(H_2)_5$+N-oscillator system. The densities are rescaled by $\sqrt{N_{rep}}$.

## S3.2 Additional results for the argon excimer

This section shows additional results for the $Ar_2$ excimer under collective strong coupling.

In Figure S3, we show the (unnormalized) squared wave function $\chi_\nu^2(R)$ for the vibrational levels $\nu = 6,7$ of the E1 PES computed at QED-CCSD-1-1 level with $g = 0.04$ a.u. (solid black lines), compared to the electronic GS and S1 levels (dotted orange lines). The results

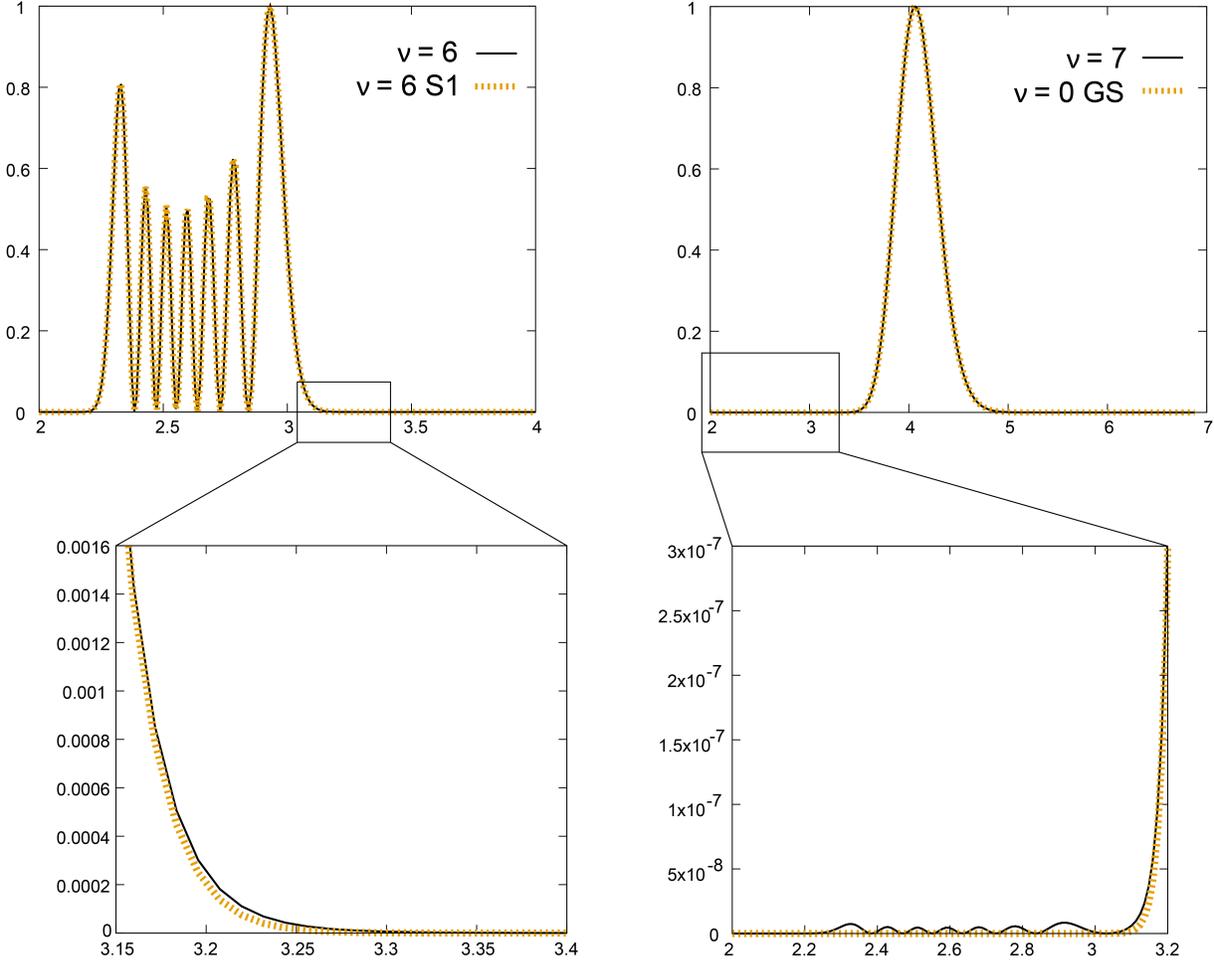

Figure S3: Unnormalized squared vibrational wave function $\chi_\nu^2(R)$ $\nu = 6,7$ for the excited state PES E1 computed at QED-CCSD-1-1 level with $g = 0.04$ a.u. compared to the electronic ground and excited state levels.

show that the levels $\nu \leq 6$ resemble the S1 vibrational states, while for $\nu \geq 7$ the levels resemble the ground state vibrations. The highlight of Figure S3 shows that for $\nu \geq 7$, the wave function has an oscillatory behaviour also in the region $2.2 - 3.0$ Å, which corresponds

to the excited state minimum.

### S3.3  Angular conical intersection in Ar$_2$

In Figure S4, we show the QED-CCSD-1-1 potential energy surfaces of the first two excited states of Ar$_2$ with field polarization perpendicular to the bond, $\lambda_0 = 0.0001$ a.u. and $\lambda_0\sqrt{N} = 0.04$ a.u. as in the main text. Contrary to the parallel-bond polarization, the avoided crossing is now an intersection. From Figure S4, it is already clear that the intersection is conical.

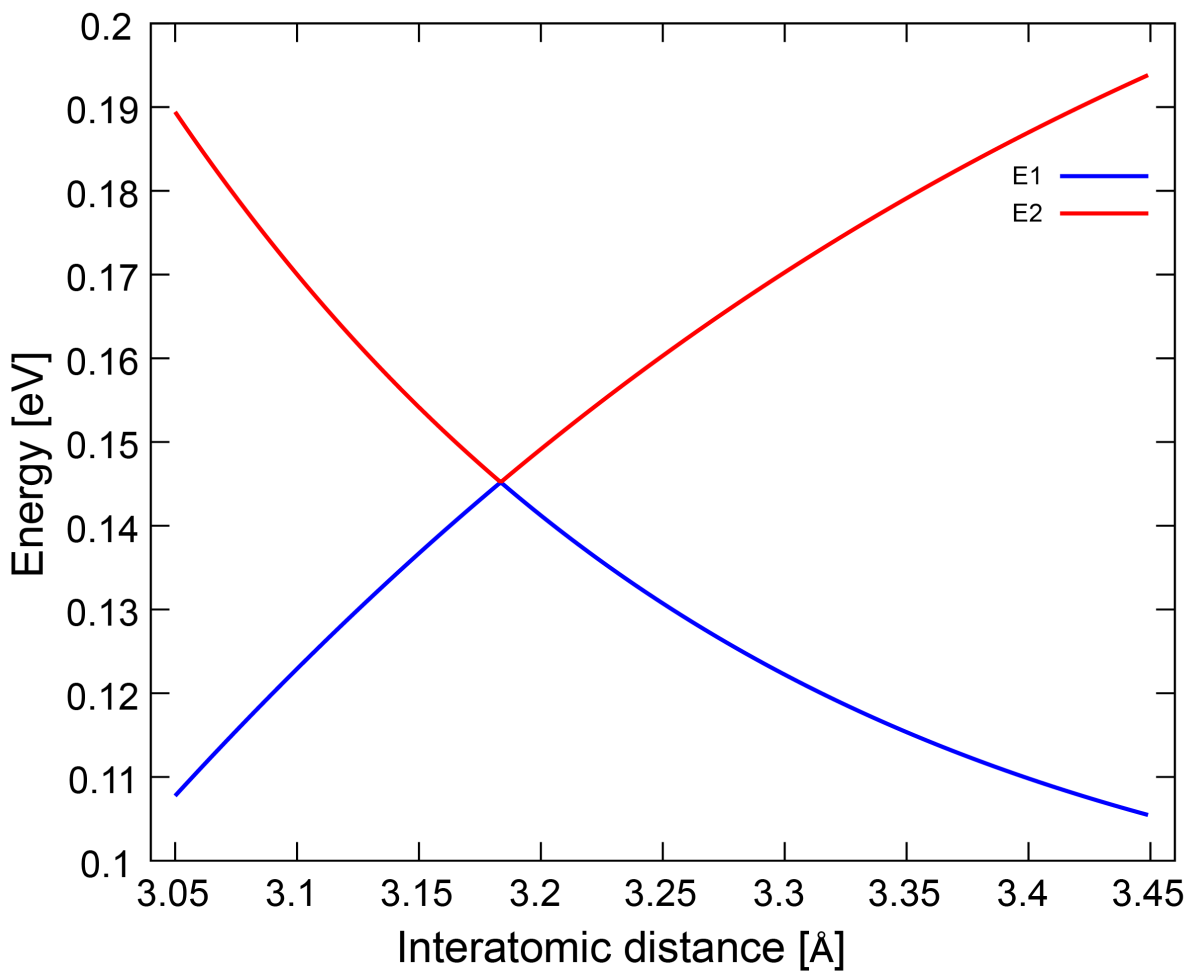

Figure S4: Potential energy surface of the first two excited states of Ar$_2$ computed using QED-CCSD-1-1 aug-cc-pVDZ with field polarization perpendicular to the bond. When the field is parallel to the bond we have an avoided crossing, while here we have an intersection.

To further confirm it, we exploit the fact that the wave function changes sign when moving

around a conical intersection[9]. To this end, rather than the CC method, we resort to the collective CIS method outlined in subsubsection S1.2.1. The system is studied using the aug-cc-pVDZ basis set, and we computed the excited state PES to locate the intersection. We then move in the space $(R, \theta)$ where $R$ is the interatomic distance, and $\theta$ is the angle between the bond and the field polarization, moving around the intersection. In Figure S5, we report the CIS wave function coefficients along a close path circumventing the intersection. The wave coefficients (and thus the wave function) change sign circumventing the intersection, which is thus a conical intersection. The light-matter coupling strength thus introduces a

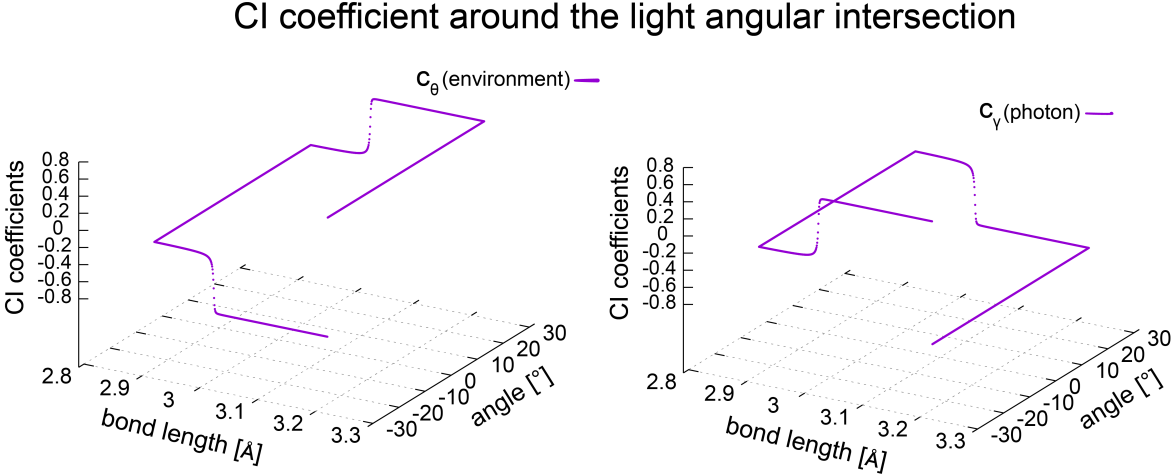

Figure S5: Photon and oscillator CI coefficients moving around the light-angular conical intersection computed at collective QED-HF/CIS level of theory. The wave function changes sign, which is an indication of a conical intersection.

novel angular conical intersection.

# S4 Molecular geometries

In this section, we report the molecular geometries of all the systems studied in the paper.

**Table S1:** Molecular geometry of the $(H_2)_5 + (H_2)_3$ system. The coordinates are provided in Angstrom Å.

|    | Atom | x [Å] | y [Å] | z [Å] |
|----|------|-------|-------|-------|
| 1  | H    | 0.39  | 0.00  | 0.00  |
| 2  | H    | -0.39 | 0.00  | 0.00  |
| 3  | H    | 0.38  | 5.0   | 0.00  |
| 4  | H    | -0.38 | 5.0   | 0.00  |
| 5  | H    | 0.38  | -5.0  | 0.00  |
| 6  | H    | -0.38 | -5.0  | 0.00  |
| 7  | H    | 0.38  | 10.0  | 0.00  |
| 8  | H    | -0.38 | 10.0  | 0.00  |
| 9  | H    | 0.38  | -10.0 | 0.00  |
| 10 | H    | -0.38 | -10.0 | 0.00  |
| 11 | H    | 0.38  | 30.0  | 0.00  |
| 12 | H    | -0.38 | 30.0  | 0.00  |
| 13 | H    | 0.38  | -30.0 | 0.00  |
| 14 | H    | -0.38 | -30.0 | 0.00  |
| 15 | H    | 0.38  | 50.0  | 0.00  |
| 16 | H    | -0.38 | 50.0  | 0.00  |

**Table S2:** Molecular geometry of the electronic CCSD ground state $Ar_2$ minimum. The coordinates are provided in Angstrom Å.

|   | Atom | x [Å] | y [Å] | z [Å] |
|---|------|-------|-------|-------|
| 1 | Ar   | 0.0   | 0.00  | 0.00  |
| 2 | Ar   | 4.075 | 0.00  | 0.00  |

**Table S3:** Molecular geometry of $Ar_2$ in the electronic CCSD excited state S1 energy minimum. The coordinates are provided in Angstrom Å.

|   | Atom | x [Å]  | y [Å] | z [Å] |
|---|------|--------|-------|-------|
| 1 | Ar   | 0.0    | 0.00  | 0.00  |
| 2 | Ar   | 2.5183 | 0.00  | 0.00  |

Table S4: Molecular geometry of Ar$_2$ at the conical intersection computed at collective QED-HF/CIS level of theory (the field polarization is along the x axis). The coordinates are provided in Angstrom Å.

|   | Atom | x [Å] | y [Å] | z [Å] |
|---|------|-------|-------|-------|
| 1 | Ar   | 0.0   | 0.00  | 0.00  |
| 2 | Ar   | 0.0   | 3.0219 | 0.00 |


\end{widetext}

\end{document}